\documentclass[preprint2,tighten]{aastex6}
\pdfoutput=1 
\usepackage{amsmath,amstext}
\usepackage{apjfonts}
\usepackage[T1]{fontenc}
\usepackage[figure,figure*]{hypcap} 
\usepackage{enumitem}
\include{hyperlink-year-only-natbib-patch}

\def\sek~{\S\,}

\newcommand*{\kms}{\,km\,s$^{-1}$}
\newcommand*{\nallnoflag}{71 }
\newcommand*{\nallnosdssflag}{202 }
\newcommand*{\nhost}{8 }
\newcommand*{\mmtspec}{10,723} 
\newcommand*{\aatspec}{6,340 }
\newcommand*{\imacsspec}{567 }

\newcommand*{\urlrm}[1]{\href{http://#1}{#1}}
\newcommand*{\urlrms}[1]{\href{https://#1}{#1}}
\newcommand*{\msun}{\ensuremath{\text{M}_\odot}}
\newcommand*{\vpeak}{V_\text{peak}}

\shorttitle{The SAGA Survey: Satellites Around Galactic Analogs}
\shortauthors{Geha et al.}

\begin{document}
\title{The SAGA Survey:\\ I.~Satellite Galaxy Populations Around Eight Milky Way Analogs}

\author{Marla~Geha\altaffilmark{1}}
\author{Risa~H.~Wechsler\altaffilmark{2,3}}
\author{Yao-Yuan~Mao\altaffilmark{4}}
\author{Erik~J.~Tollerud\altaffilmark{5}}
\author{Benjamin~Weiner\altaffilmark{6}}
\author{Rebecca~Bernstein\altaffilmark{7}}
\author{Ben~Hoyle\altaffilmark{8,9}}
\author{Sebastian~Marchi\altaffilmark{10}}
\author{Phil~J.~Marshall\altaffilmark{3}}
\author{Ricardo~Mu\~noz\altaffilmark{10}}
\and
\author{Yu~Lu\altaffilmark{7}}

\altaffiltext{1}{Department of Astronomy, Yale University, New Haven, CT 06520, USA}
\altaffiltext{2}{Kavli Institute for Particle Astrophysics and Cosmology \& Department of Physics, Stanford University, Stanford, CA 94305, USA}
\altaffiltext{3}{SLAC National Accelerator Laboratory, Menlo Park, CA 94025, USA}
\altaffiltext{4}{Department of Physics and Astronomy \& Pittsburgh Particle Physics, Astrophysics and Cosmology Center (PITT PACC), \\ University of Pittsburgh, Pittsburgh, PA 15260, USA}
\altaffiltext{5}{Space Telescope Science Institute, 3700 San Martin Dr, Baltimore, MD 21218, USA}
\altaffiltext{6}{Department of Astronomy, University of Arizona, Tucson, AZ, USA}
\altaffiltext{7}{The Observatories of the Carnegie Institution for Science, 813 Santa Barbara St., Pasadena, CA 91101, USA}
\altaffiltext{8}{Universitaets-Sternwarte, Fakultaet f\"ur Physik, Ludwig-Maximilians Universitaet Muenchen, Scheinerstr.\ 1, D-81679 Muenchen, Germany}
\altaffiltext{9}{Max Planck Institute f\"ur Extraterrestrial Physics, Giessenbachstr.\ 1, D-85748 Garching, Germany}
\altaffiltext{10}{Departamento de Astronomia, Universidad de Chile, Camino del Observatorio 1515, Las Condes, Santiago, Chile}

\begin{abstract}
We present the survey strategy and early results of the ``Satellites Around Galactic Analogs'' (SAGA) Survey.  
  The SAGA~Survey's goal is to measure the distribution of satellite  galaxies around 100 systems analogous to the Milky Way down to the  luminosity of the Leo I dwarf galaxy ($M_r<-12.3$).  We define a Milky Way  analog based on  $K$-band luminosity and local environment.  Here, we
  present satellite luminosity functions for \nhost Milky Way analog galaxies between 20 to 40~\,Mpc.  These systems have nearly complete spectroscopic coverage of candidate satellites within the projected host virial radius down to $r_o<20.75$ using low redshift $gri$ color criteria.  
We have discovered a total of 25 new satellite galaxies:  14~new satellite galaxies meet our formal criteria around our complete host systems, plus 11 additional satellites in either incompletely surveyed hosts or below our formal magnitude limit. Combined with 13 previously known satellites, there are a total of 27 satellites around \nhost complete Milky Way analog hosts.
We find a wide distribution in the number of satellites per host, from 1 to 9, in the luminosity range for which there are five Milky Way satellites.  
Standard abundance matching extrapolated from higher luminosities predicts less scatter between hosts and a steeper luminosity function slope than observed.
We find that the majority of satellites (26 of 27) are star-forming.   These early results indicate that the Milky Way has a different satellite population than typical in our sample, potentially changing the physical interpretation of measurements based only on the Milky Way's satellite galaxies.
\end{abstract}

\section{Introduction}
The Milky Way is the most well-studied galaxy in the Universe \citep[e.g.,][]{Bln2016}.   From a cosmological and galaxy formation perspective, one of the more informative components of the Milky Way is its population of
dwarf galaxy satellites.  While the number of faint satellites ($M_r > -10$) is steadily increasing due to discoveries in on-going large-area imaging surveys \citep[e.g.,][]{2015ApJ...805..130K, 2015ApJ...813..109D}, the number of bright satellites ($M_r < -10$) has remained unchanged since the discovery of the disrupting Sagittarius dwarf spheroidal galaxy over 20 years ago \citep{1994Natur.370..194I}.   The population of bright satellite galaxies around the Milky Way is thus largely complete.

The properties of the Milky Way's brightest satellites do not agree with predictions of the simplest galaxy formation models based on simulations of the Lambda Cold Dark Matter model ($\Lambda$CDM).  $\Lambda$CDM simulations including only dark matter, combined with simple galaxy formation prescriptions, over-predict both the number of satellite galaxies observed around the Milky Way and their central mass densities \citep[the "too-big-to-fail" problem;][]{BK2012a}.  Stated differently, $\Lambda$CDM predicts large numbers of dark matter subhalos that either do not exist around the
Milky Way, do not host bright satellite galaxies, or are not as dense
as expected \citep[e.g.,][]{GK2014}.  
It has been suggested that a realistic treatment of baryonic physics and the stochastic nature of star formation can fix these discrepancies \citep[e.g.,][]{Brooks2014,Guo2015,Wetzel2016,Brooks2017}. 
Other authors have suggested that the
discrepancy favors alternative models to CDM \citep[e.g.,][]{Lovell2012,Polisensky2014}.  
While some of these discrepancies may be solved if the Milky Way has a lower mass \citep[e.g.,][]{2013MNRAS.428.1696V,dierickx17},  similar results hold for M31 in the Local Group \citep{Tollerud2014}. 

It is possible that the Local Group satellites are not representative of typical galaxies at this mass scale \citep[e.g.,][]{Purcell2012, Jiang2016}.
Several studies have considered the question of how typical the Milky Way is in terms of its bright satellite population \citep{Liu2011, Guo2011, Tollerud2011, james2011,Strigari2012,robotham2012}.   Most of these studies use the Sloan Digital Sky Survey (SDSS) whose spectroscopic magnitude limit of $r = 17.7$ corresponds to satellites in the nearby Universe that are similar to the Large and Small Magellanic Clouds ($M_r = -18.6$ and $-17.2$, respectively).  These studies find that our Galaxy is unusual, but not
yet uncomfortably so, in its bright satellite population. The Milky Way analogs on average have only 0.3 satellites brighter than
these luminosities, versus two for the Milky Way \citep{Liu2011}. The
distribution of these bright satellites is also remarkably consistent with simulations using fairly straightforward assumptions about the galaxy-halo connection \citep{Busha2011, 2013ApJ...773..172R,Kang2016}. It is below these luminosities that the Milky Way's satellite properties diverge from simple galaxy formation predictions. This suggests a search for such satellites around a large sample of hosts.

Identifying satellites fainter than the Magellanic Clouds in a statistical sample of host galaxies is observationally challenging.  For hosts within 10\,Mpc, satellites can be reliably distinguished based on size, surface brightness, and, in the most nearby cases, by resolved stars \citep[e.g.,][]{pisces2016,Javanmardi2016,Danieli17}.   However, this volume contains only a handful of Milky Way-like galaxies, and is thus insufficient to answer the statistical questions outlined above.  Low mass galaxies around  Milky Way analogs beyond 10\,Mpc are difficult to distinguish from the far more numerous background galaxy population using photometry alone.   Previous studies have focused on spectroscopic follow-up of single hosts \citep{Spencer2012}, or attempted to constrain satellite populations statistically \citep{Speller2014}.  Photometric redshifts do not perform well at low redshifts (see \autoref{ssec_photz}), thus a wide-area spectroscopic survey is required to 
quantify satellite populations.  Currently available spectroscopic surveys deeper than the SDSS either cover fairly small areas, are too shallow \citep[e.g., GAMA, $r < 19.8$;][]{2015MNRAS.452.2087L} and/or use color cuts specifically aimed at removing low redshift galaxies \citep[e.g., DEEP2; ][]{DEEP2}.

To investigate the current small-scale challenges in $\Lambda$CDM requires characterizing the complete satellite luminosity function at least down to $M_r \sim -12$ ($M_{\rm star}\sim 10^6 \, \msun$). At this scale, current galaxy formation models based on CDM halos fail to reproduce the luminosity and velocity functions of observed galaxies. In addition, to differentiate between solutions (e.g., baryonic feedback and alternative dark matter) requires a better understanding of the host-to-host scatter in the satellite luminosity function,
which in turn requires identifying luminosity functions for many tens of hosts.

In this paper, we present the survey strategy and early results of the Satellites Around Galactic Analogs (SAGA) Survey\footnote{\urlrm{sagasurvey.org}\label{footnote:sagawebsite}}.   The goal of the SAGA~Survey is to obtain spectroscopically-confirmed complete satellite luminosity functions within the viral radius of 100 Milky Way analogs in the distance range $20 - 40$\,Mpc down to $M_r=-12.3$.   The paper is organized as follows.  In \autoref{sec_survey}, we describe the SAGA~Survey strategy including our definition and selection of Milky Way host analogs.  In \autoref{sec_targeting}, we detail the observing facilities used to obtain redshifts for over 17,000 candidate satellite galaxies.  In \autoref{sec_selection} and \autoref{sec_incomplete}, we describe our efforts to improve targeting efficiency for low redshift galaxies and explore possible biases in our existing redshift survey.   Finally, in \autoref{sec_results} we present results based on satellites discovered around eight Milky Way host galaxies.

All distance-dependent parameters in this paper are calculated
assuming $H_0 = 70$\,\kms\,Mpc$^{-1}$.  Magnitudes and colors are
extinction corrected (as denoted with a subscript 'o', e.g., $r_o$) using \citet{schlegel98} as reported by SDSS DR12
and $K$-corrected to redshift zero using the {\tt kcorrect v4\_2}
software package \citep{kcorrect07}.

\begin{figure*}[htb!]
\centering
 \includegraphics[width=0.645\textwidth]{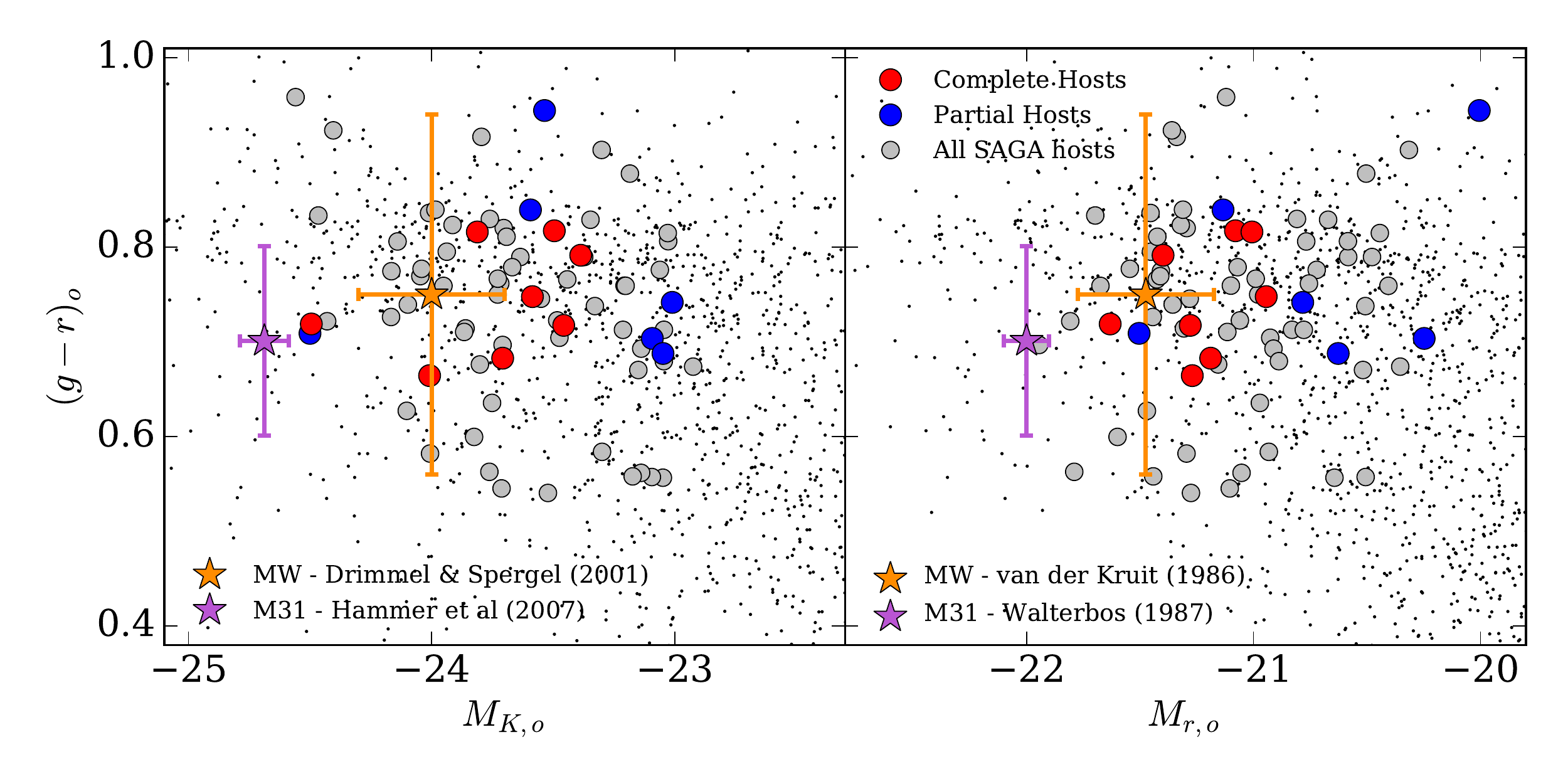}
 \includegraphics[width=0.35\textwidth]{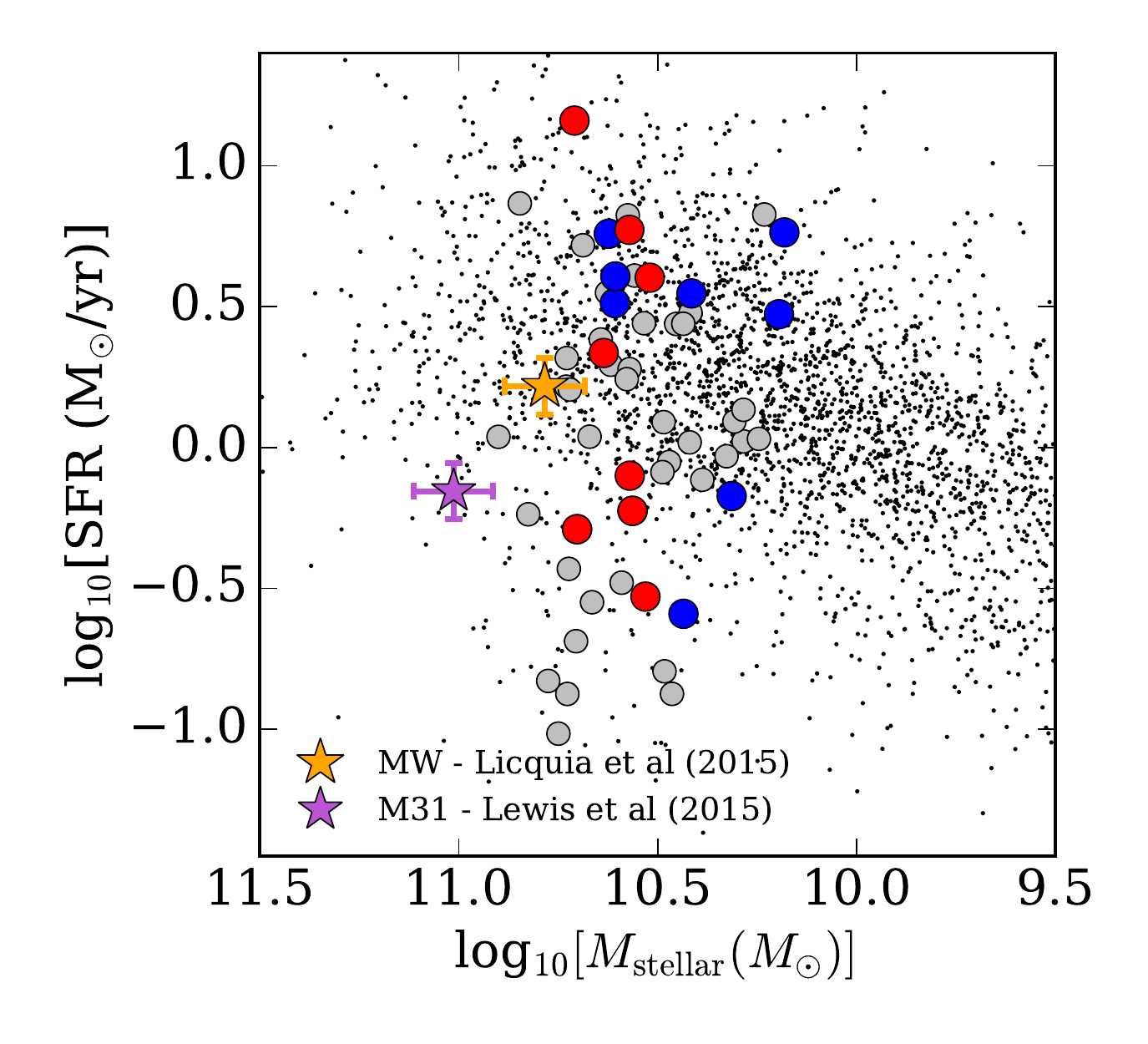}
\vskip -0.2cm
\caption{We define our Milky Way analog sample based on the total $K$-band luminosity and local environment.   Absolute magnitude $M_{K,o}$ ({\it left}) and $M_{r,o}$ ({\it center}) is plotted versus $(g-r)_o$ color for all galaxies within 40\,Mpc
  (black dots).  We plot galaxies which pass our Milky Way analog criteria as grey circles, and the analogs presented in this paper in red (blue) circles which have completed (partial) spectroscopic coverage.  ({\it
    Right})  Stellar mass versus star formation rate for the same sample.    
     The Milky Way itself is shown in each panel as the orange star, M31 is shown as a purple star.   Rather than perfectly replicating the Milky Way itself, our definition is designed to facilitate matching to simulated systems which are representative of galaxies similar to the Milky Way (see \autoref{fig_MK_Mhalo}).
   \label{fig_hosts}}
\end{figure*}
 
\section{The SAGA Survey Design}\label{sec_survey}

The goal of the SAGA survey is to characterize the satellite
galaxy population around 100 Milky Way analogs within the virial radius down to an absolute
magnitude of $M_{r,o} = -12.3$.   In the Milky Way, there are five satellites brighter than this magnitude limit; the dimmest of these, Leo I, has a luminosity of $M_{r,o} = -12.3$ and a stellar mass of $M_* = 3\times 10^6 \, \msun$ \citep{2012AJ....144....4M}.  We chose Milky Way analog galaxies from a largely complete list of galaxies in
the local universe (\autoref{ssec_master}), and describe a well-defined set of criteria for selecting Milky
Way analog galaxies (\autoref{ssec_defineMW}).  We first motivate our Galactic analog  selection over the distance range 20 to 40\,Mpc, and then simulate the properties of Milky Way
satellites observed at these distances (\autoref{ssec_mwsats}).

\subsection{The Master List: A Complete Galaxy Catalog in a 40\,Mpc Volume}\label{ssec_master}
 
We select Milky Way analog galaxies from a catalog of galaxies that is as complete as possible within the survey volume in order to
better understand our selection function.  To do this requires a catalog of all bright galaxies in the Local Universe.  
We found that no single available catalog provided an adequate sample and therefore compiled this Master List ourselves.  

The Master List is a complete catalog of all galaxies within $v <
3000$ \kms\ brighter than $M_K = -19.6$.  The completeness
limit is set by the magnitude limit of the 2MASS redshift survey
\citep[$K_s<13.5$ mag;][]{2000AJ....119.2498J}.  Our master list includes fainter galaxies and galaxies out to 4000 \kms, but this
portion of the catalog is not complete.  The catalog is primarily
based on the Hyper-LEDA2 database \citep{HyperLeda}, however,
Hyper-LEDA is missing a small fraction of galaxies in our target
volume and does not contain all physical properties required. We
therefore supplement HyperLEDA with data from various sources
including the Nearby Galaxy Catalog \citep{Karachentsev2013}, the
2MASS redshift survey and Extended Source Catalogs
\citep{2000AJ....119.2498J}, the NASA--Sloan Atlas
\citep[NSA;][]{NSA}, and the 6dF redshift survey
\citep{2009MNRAS.399..683J}.   All magnitudes are extinction corrected and $K$-corrected to redshift zero.   

\subsection{The SAGA Milky Way Analog Sample}\label{ssec_defineMW}

\begin{figure}[htb!]
\centering\includegraphics[width=\columnwidth]{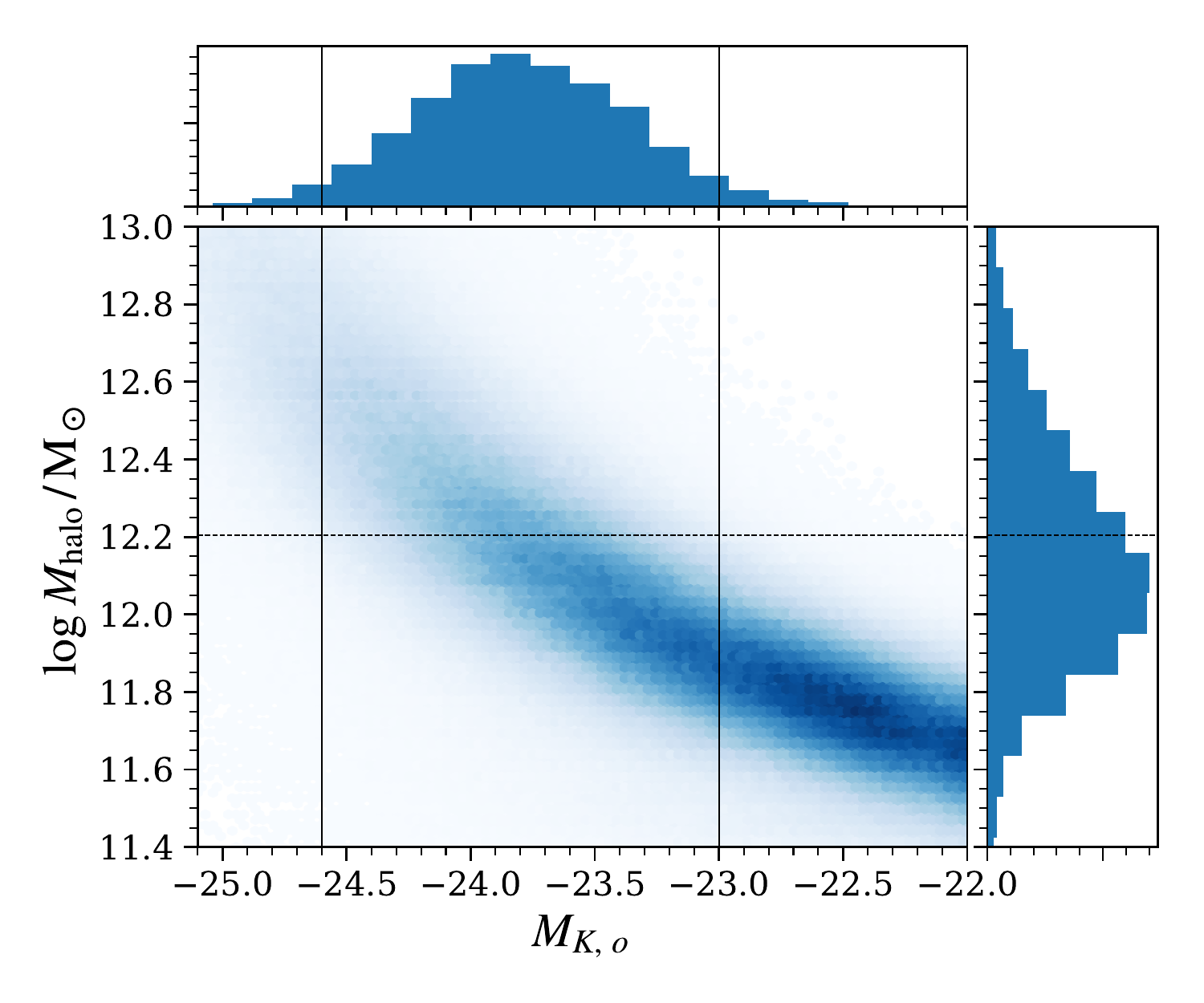}
\caption{To determine the $K$-band magnitude range over which we select Milky Way analogs, 
we use the abundance matching technique.  (\textit{Main panel}) Joint distribution (in linear color scale) of $K$-band magnitude and halo mass for isolated distinct halos.
The horizontal dashed line shows a halo mass of $1.6 \times 10^{12} \,\msun$.
The solid vertical lines show the magnitude range, $-23 > M_K > -24.6$, over which this halo mass can plausibly host a Milky Way analog.  (\textit{Top}) Conditional distribution of $K$-band magnitude at a fixed halo mass of $1.6 \times 10^{12} \,\msun$.   (\textit{Right}) Conditional distribution of halo mass, integrated over our $K$-band magnitude range, $-23 > M_K > -24.6$.
\label{fig_MK_Mhalo} }
\end{figure}

In searching for satellites around Milky Way analogs, a key question
is how to best define the Milky Way itself.  The observed properties of the Milky Way are uncertain and vary significantly between published measurements \citep[e.g.,][]{Bln2016,2015ApJ...809...96L}.   Furthermore, many of our science questions are best answered by comparing hosts with similar dark matter halo masses, a property which is impossible to directly measure at this mass scale.    Rather than perfectly replicate the Milky Way itself, our definition is designed to facilitate matching the observed analog sample to simulated systems which are representative of galaxies similar to the Milky Way.  


We define the Milky Way based on its total $K$-band luminosity and local environment.
We chose the $K$-band luminosity as a simple proxy for a stellar mass.  We first assume a dark matter halo mass of the Milky Way and then use the abundance matching technique \citep[see e.g.,][]{Kravtsov2004,Vale2004,Vale2006,Conroy2006,Behroozi2010} to obtain the corresponding $K$-band luminosity range.
We assume that the Milky Way halo mass is $1.6 \times 10^{12} \,\msun$. This choice is consistent with various estimates of the Milky Way halo mass which range from $0.6 - 2.7 \times 10^{12}\, \msun$ \citep[for a summary of this literature, see Table~8 of][]{Bln2016}.
We then use a publicly available abundance matching code\footnote{\urlrms{bitbucket.org/yymao/abundancematching}} to match the number density of halo maximal circular velocity at its peak value on the halo's main branch (commonly known as $\vpeak$) to the $K$-band luminosity function from the 6dF Galaxy Survey \citep{jones2006}.
The halo $\vpeak$ function is extracted from a dark matter-only simulation, which has a side length of 250 Mpc$h^{-1}$ and 2560$^3$ particles, with cosmology parameters and simulation code identical to those in the Dark Sky Simulations \citep{skillman2014}. 
We assume a 0.15\,dex scatter in luminosity at fixed halo $\vpeak$ \citep[see e.g.,][for current constraints on the scatter]{2013ApJ...771...30R,Gu2016,Lehmann2017}.  

\autoref{fig_MK_Mhalo} shows the galaxy--halo connection described here and demonstrates the large scatter between halo mass and $M_K$. 
With this method, we can obtain the distribution of $M_K$ of isolated distinct halos at any given halo mass. For a halo mass of $1.6 \times 10^{12} \,\msun$ (horizontal dashed line in \autoref{fig_MK_Mhalo}), the 95\% interval is $-23 > M_K > -24.6$, as shown in the top panel of \autoref{fig_MK_Mhalo}.
This range encompasses the range of $M_K$ reported for the Milky Way in the literature \citep{1996ApJ...473..687M,2001ApJ...556..181D,2002ApJ...573..597K}, but is slightly larger due to the scatter between halo mass and $\vpeak$ (i.e., the scatter in halo concentration at a fixed halo mass) and also the assumed scatter in the galaxy--halo connection.

To match the Milky Way's large-scale environment, we impose an isolation criterion on our analog hosts such that there are no galaxies brighter than $M_K +1$ within one degree of the host, and such that the host is not within 2 virial radii of a
massive ($5\times10^{12}\,\msun$) galaxy in the 2MASS group
catalog \citep{2011MNRAS.416.2840L}.  The former isolation criteria is much more restrictive than the latter, together the environment cuts reduce the number of Milky Way-like galaxies by a factor of two.  These criteria are agnostic to the presence of a M\,31-like companion, as the Milky Way is slightly beyond 2 virial radii from M\,31.   19 of our \nallnoflag Milky Way analogs have nearby M\,31-mass galaxies between 0.8 to 2 Mpc, while none of the hosts listed in \autoref{table_hosts} have an M\,31-mass galaxy within 2\,Mpc in 3D redshift space.  We will investigate the influence of a M\,31 companion on satellite distributions as our survey size increases, although numerical simulations suggest this may be a small affect on the subhalo population \citep{2014MNRAS.438.2578G}.

To match our available observing resources, we require our SAGA Milky Way analogs to lie in the distance range 20 to 40\,Mpc ($0.005 < z < 0.01$).   We assume the virial radius of a typical Milky Way is 300\,kpc, corresponding to the virial radius of a halo whose virial mass is $1.6\times10^{12}\,\msun$ in our assumed cosmology.  Our lower distance
bound of 20\,Mpc is set to ensure that this physical virial radius is less than $1^{\circ}$\,on the sky.  This angular radius matches the field-of-view of available spectroscopic
instrumentation (\autoref{sssec_data}) and allows efficient follow-up to confirm satellite
systems.  We note that there are fewer than 20 Milky Way analogs passing our other criteria inside 20\,Mpc, so these would not provide sufficient statistics on their own.  The upper distance bound of 40\,Mpc corresponds to the furthest distance at which a satellite similar to the Leo~I dSph ($M_r = -12.3$) corresponds to an apparent magnitude brighter than $r_o < 20.75$, a magnitude range that is both well measured in SDSS photometry and easily accessible by our
spectroscopic follow-up facilities.  We also require systems with galactic latitude
$b > 25^\circ$\, from the Galactic Plane in order to avoid fields with very high stellar foregrounds.

\begin{figure*}[t!]
\centering\includegraphics[width=0.85\textwidth]{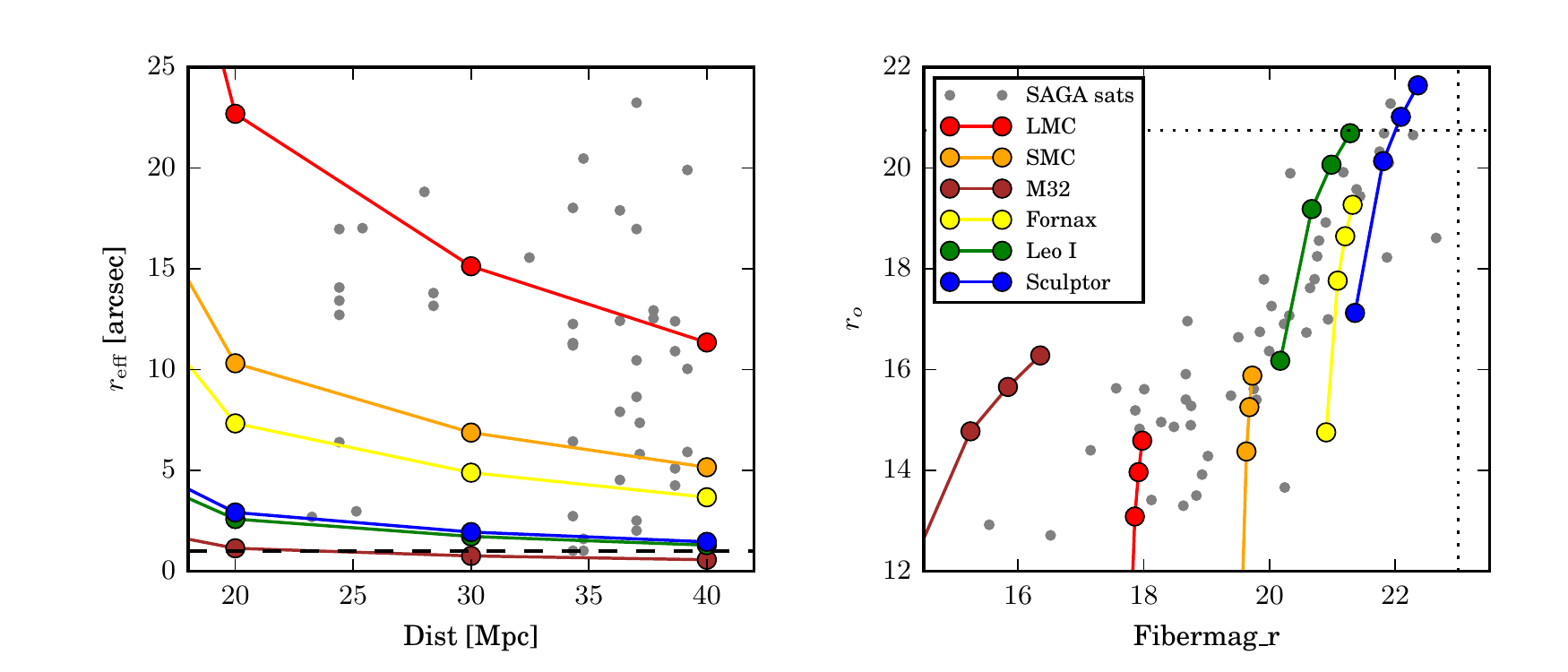}
\vskip -0.2cm
\caption{Simulated distance-dependent properties of bright Milky Way
  satellite galaxies (red through blue circles), and the M\,31 satellite M\,32 (brown circle).  In both
  panels, grey points are SAGA-identified satellites around the 16 Milky Way
  hosts listed in \autoref{table_hosts}. In all panels, symbols are plotted at the inner edge, middle and outer edge of our survey volume (20, 30 and 40\,Mpc).  ({\it Left\/}) Effective $r$-band radius as a function of
  distance.  All bright Milky Way satellites are resolved ($>1.5\arcsec$) in SDSS photometry in the SAGA~Survey volume.   Galaxies similar to the rare compact elliptical M\,32 are unresolved in SDSS and are not included in this survey. 
  ({\it Right}) Magnitude within a $1.5\arcsec$ fiber versus
  apparent $r$ magnitude.  The SAGA~Survey limits, $r_O < 20.75$ and $\text{\tt
    FIBERMAG\_R} < 23$, are plotted as dotted lines.     Galaxies as faint as Sculptor ($M_r = -11.3$) cannot be detected over our full volume, while galaxies similar to Leo~I ($M_r = -12.3$) are detectable throughout the SAGA~Survey volume. \label{fig_mwsims}}
\end{figure*}

There are \nallnosdssflag Milky Way-analogs passing our criteria.  We use SDSS DR12 as the targeting photometry for our spectroscopic follow-up and require photometry covering at least 90\% of the host's virial area.  While 93 hosts lie within the SDSS footprint, there are only \nallnoflag hosts passing this coverage criteria.  To achieve our goal of 100 systems, we plan to expand our survey outside the SDSS footprint using deeper publicly available imaging.  In this paper, we present results for \nhost hosts for which we have obtained nearly complete spectroscopy in our defined criteria using SDSS targeting photometry, as well as incomplete results for an additional eight hosts.  Details for these hosts can be found in
\autoref{table_hosts}.

We compare in \autoref{fig_hosts} the distribution of observed $M_{K,o}$, $M_{r,o}$ and $(g-r)_o$ for our Milky Way analog sample  relative to the other galaxies, the Milky Way and M\,31.  The \nhost hosts presented in this paper are shown as red symbols, blue symbols are hosts for which we have partial spectroscopic coverage and grey symbols are all \nallnoflag SAGA analog galaxies.   For the Milky Way, we assume the properties shown in \autoref{fig_hosts} from \citet{2001ApJ...556..181D,vanderkruit86,2015ApJ...809...96L}.   For M\,31, we assume physical properties from \citet{hammer07,Walterbos87,Lewis15,Courteau11}. These values are listed in \autoref{table_hosts}.

Our analog sample spans a range in observed properties that is larger than the uncertainty in the observed properties of the Milky Way itself due to our assumed scattered in the halo mass-luminosity relationship.   In the right panel of \autoref{fig_hosts}, we show the distribution of stellar mass versus total star formation rate.  Stellar masses are calculated using {\tt kcorrect v4\_2} which is based on SDSS colors \citep{kcorrect07}.  Star formation rates are determined from the total IRAS fluxes \citep{IRAS}, using the transformations of \citet{kewley2002}.   Since we have imposed no color criteria, our SAGA sample spans a wide range of star formation rates which includes the Milky Way and M\,31 values.  This range of star formation rates is represented in the sub-sample of hosts presented in this paper.

\subsection{Milky Way Satellite Properties around SAGA Hosts}\label{ssec_mwsats}

As detailed in \autoref{sec_incomplete}, interpreting the SAGA results requires an understanding of our survey completeness relative to the expected Milky Way satellite population.  To build this understanding, we simulate the expected properties of Milky Way and Local Group
 satellites at the distances of our survey host galaxies ($20-40$\,Mpc).  There are
currently over 50 confirmed or candidate satellite galaxies in orbit
around the Milky Way \citep[e.g.,][]{2015ApJ...813..109D}, although
this number is based on incomplete sky coverage.  However, at the distances of our SAGA sample, we can detect only analogs of the brightest  Milky
Way satellites where the Milky Way census is likely complete.  An
apparent magnitude of $r_o=20.75$ (the maximum extinction-corrected
spectroscopic depth of our survey) corresponds to $M_r = -12.3$ at the
outer survey limit of 40\,Mpc, and $M_r = -10.8$ at the inner limit of
20\,Mpc.  In the Milky Way, there are 5 satellites down to
$M_r = -12.3$ (the Large/Small Magellanic Clouds, the disrupting
Sagittarius dSph, Fornax and Leo~I dSph), and 6 satellites down to
$M_r = -10.8$ (plus Sculptor).  Throughout this
paper, we distinguish between satellite statistics for the full survey and those for which we
are complete throughout our survey volume ($M_r < -12.3$).

In \autoref{fig_mwsims}, we simulate basic properties of the Milky
Way satellites at the distances of the SAGA hosts.  We use properties of the Milky Way satellites from
\citet{2012AJ....144....4M}, supplementing the size and color of the Magellanic Clouds from \citet{bothun88}.   We calculate the effective $r$-band radius and apparent $r$-band magnitude by shifting the observed quantities without cosmological correction.   We calculate the apparent magnitude within the $3\arcsec$ diameter SDSS fiber ({\tt fibermag\_r} in the SDSS database) using the total magnitude of each satellite and assuming an exponential light profile. 

Satellite galaxies analogous to the Fornax dSph ($M_r = -13.7$) are well detected
throughout our survey volume. Sculptor-like satellites ($M_r =
-11.0$) fall below our magnitude limit in the outer half of our survey volume. The SAGA~Survey's goal is to detect down to Leo~I-like galaxies. At the outer limit of our survey
volume, we can just detect a Leo\,I-like satellite ($M_r = -12.3$),
although this object is marginally resolved ($1.5\arcsec$) in the SDSS-data at the
outer edge of our survey.   We conclude that using SDSS photometry, Leo~I analogs are detectable throughout the SAGA survey volume.

We include in \autoref{fig_mwsims} the compact elliptical galaxy
M\,32 as brown points.   As shown in the right panel,
M\,32 would appear unresolved (less than $1\arcsec$) throughout our survey volume in the SDSS photometry and would likely be classified as a star.    At the beginning of our survey, we followed-up stars in the magnitude range $19.5 <r_o < 20.75$, but chose not to continue follow-up of unresolved  objects in the main SAGA~Survey.   This choice is discussed further in  \autoref{ssec_m32}.   Our luminosity functions are therefore biased against
compact M\,32-like satellites.

\section{The Data}\label{sec_targeting}

We next describe our use of SDSS Data Release 12 (DR12) imaging \citep{DR12} to target
candidate satellite galaxies (\autoref{sssec_sdssphot}).   We then describe the telescope
facilities used to obtain follow-up spectroscopy to measure redshifts of these targets (\autoref{sssec_data}).

\subsection{SDSS Photometry}\label{sssec_sdssphot}

For each Milky Way analog (host) galaxy, we select all SDSS DR12 photometric objects
within a one degree radius.  We then clean these catalogs using the criteria described below to remove spurious objects, while avoiding removal of real galaxies.  The two major contaminants in the SDSS photometric catalogs are shredded parts of nearby extended galaxies and very faint targets that are measured to be bright due to poor sky subtraction.   We address these issues below.

We first remove targets with bad
photometry as flagged by the SDSS {\it Photo} pipeline using the {\tt SATURATED, BAD\_ERROR}
and {\tt BINNED1} flags.  We then remove objects whose median photometric
error in the $gri$ bands is greater than 0.5\,mag.  We find that a more stringent cut on photometric errors removes objects of interest.  For galaxies
brighter than $r=18$, we require the Petrosian radius measured in the
$g$, $r$ and $i$-bands to agree within $40^{\prime\prime}$ to address an issue
due to poor sky subtraction.   The SDSS photometric pipeline is not optimized for large extended
objects and nearby galaxies are often split into many fainter photometric objects (HII regions, spiral arms, etc.) instead of being a single object.  To address the issue of photometric shredding, we use the NASA-Sloan Atlas (NSA) version 0.1.3 \citep{NSA}, which is a reprocessing of
the SDSS photometry for galaxies with $z < 0.055$ using an improved
background subtraction technique.  For each galaxy in the NSA, we
replace the SDSS DR12 photometric properties with the NSA measurements
and remove all other objects within twice the elliptical NSA-measured
$r_{90}$ radius.  For galaxies with SDSS spectroscopy beyond
$z > 0.055$, we use the measured SDSS DR12 radius to remove objects
within a more conservative region of $r_{90}$.  Any photometric objects identified within 10\,kpc in projection of the main host are assumed to be part of the host and flagged as such.  In addition, we do a visual inspection of all targets.  We remove by hand a small number of objects that were not caught by our automated criteria, and add back in a handful of objects which were erroneously flagged as poor, usually faint galaxies near bright stars.  For each galaxy, we use the fitted exponential light profile to calculate an effective $r$-band surface brightness, $\mu_{r,\rm eff}$, which is the average surface brightness inside the effective half-light radius.

  For our main target selection, we concentrate only on objects brighter
  than an extinction-corrected $r_o < 20.75$ and select objects classified by SDSS as galaxies
  using the SDSS star/galaxy criteria ($\text{\tt type} = 3$).  We note that the maximum variation in foreground $r$-band extinction values across any individual host is between 0.03 - 0.07 magnitudes.
   For a discussion on star-like objects in the survey, see \autoref{ssec_m32}.  We require the
  magnitude measured within the SDSS fiber to be brighter than $\text{\tt FIBERMAG\_R} < 23$.
  While this requirement might remove very low surface brightness
  objects, in practice we find the criteria removes only noise
  fluctuations in the SDSS data and would not remove a Milky Way satellite at these distances (right panel, \autoref{fig_mwsims}).     We
  base our spectroscopic follow-up on these `cleaned' SDSS photometric
  catalogs.

\begin{figure*}[htb!]
\centering\includegraphics[width=0.9\textwidth]{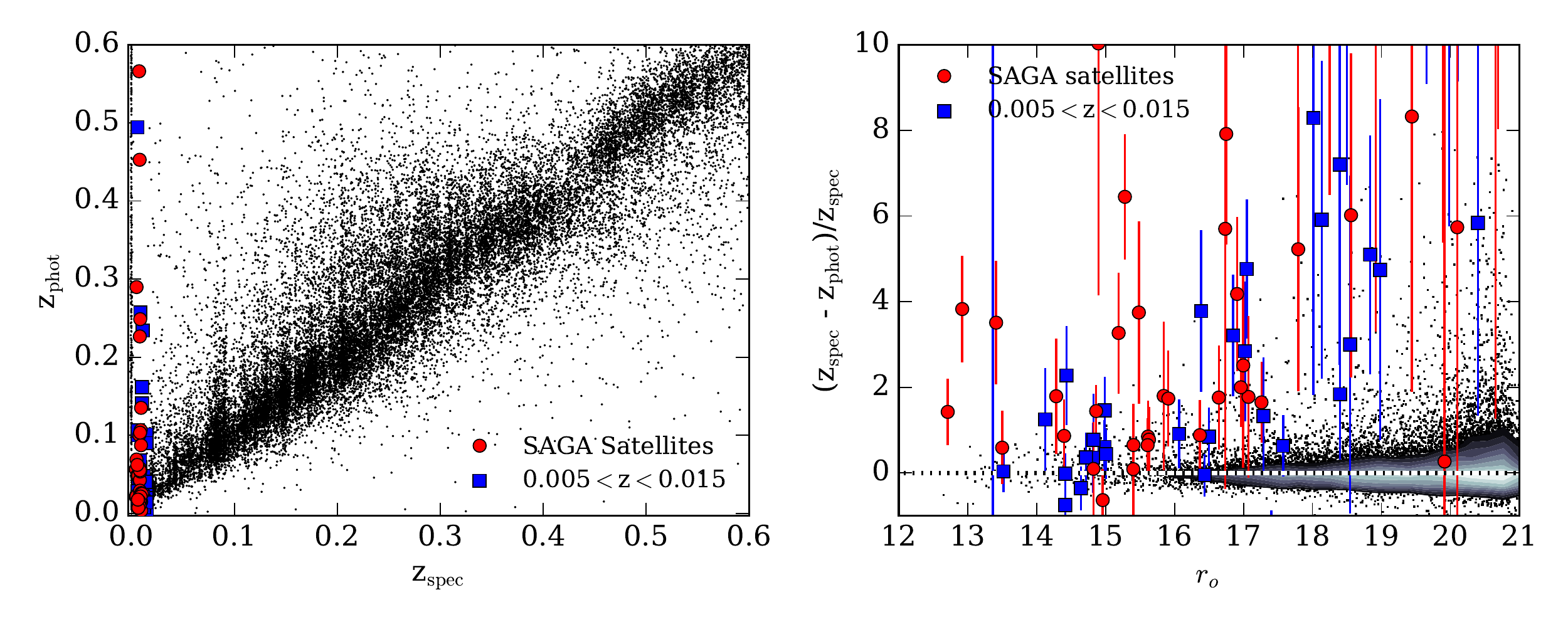}
\vskip -0.5cm
\caption{({\it Left}\/) Spectroscopic redshift plotted against the SDSS DR12 photometric redshifts from \citet{sdss_photoz16}. ({\it Right}\/)  Apparent $r$-band magnitude versus the fractional difference between the spetroscopic and photometric redshifts.  In both panels, we plot SAGA satellite galaxies as red circles and a larger number of field galaxies over a similar redshift range ($0.005 < z < 0.015$) as blue squares.   For the majority of galaxies with redshift $z < 0.015$, particularly for galaxies fainter than $r_o > 17.7$, photometric redshifts are neither accurate nor precise.\label{fig_photoz}}
\end{figure*}

\subsection{Spectroscopic Observations and Data Reduction} \label{sssec_data}

We obtained redshifts for over 17,000 objects which did not previous have redshifts in the literature.   A summary of all hosts for which we have taken spectroscopic data is shown in \autoref{table_hosts}.   These data were taken primarily with the MMT/Hectospec and
AAT/2dF systems over the period 2012--2017.  Below we briefly
describe the observational set-up and data reduction for each system.  
In most cases, we prioritized targeting galaxies inside of the host virial radius which were brighter than our survey limit of $r_o < 20.75$ and passed our $gri$ color criteria as described in \autoref{ssec_gri}.  Secondary priority was given to targets 0.5\,magnitudes fainter than our survey limit passing our $gri$ color criteria inside the virial radius.   Finally, we filled in our pointings with galaxies passing our main survey criteria, but were beyond the host virial radius.    


\begin{description}[wide=0\parindent]

\item[MMT/Hectospec] Hectospec is a fiber-fed spectrograph
on the MMT 
which deploys 300 fibers over
a one degree diameter field \citep{2005PASP..117.1411F}.  We used Hectospec with the 270 l/mm grating
resulting in wavelength cover of 3650 - 9200$\mbox{\AA}$ and $1.2
\mbox{\AA}$ per pixel and spectral resolution R$\sim 1000$.  MMT fields were designed using the Hectospec
observation planning software \citep[XFITFIBS,][]{mmt_target98}.  Exposure times
ranged between 1 to 2 hours per configuration.  The data were reduced
using the HSRED pipeline \citep{cool2008} and derived
redshifts with RVSAO \citep{mmt_rv98} with templates constructed
for this purpose.  We obtained \mmtspec\ spectra with MMT/Hectospec between May 2013 and March 2017.

\item[AAT/2dF] 2dF is a fiber-fed spectrograph on the
Anglo-Australian Telescope (AAT) with 400 fibers over a 2 degree
diameter field.  We used the 580V and 385R gratings in the blue and
red arms, respectively, both providing a resolution of $R =1300$ (between 1 to $1.6 \mbox{\AA}$ per pixel) over a
maximum wavelength range of $3700 - 8700\mbox{\AA}$.  25 fibers were
on regions of blank sky (defined as neither SDSS nor USNO-B detections
within $5^{\prime\prime}$).  Data were reduced using the facility software {\tt 2dfdr}, {\tt autoz}
\citep{2014MNRAS.441.2440B} and {\tt marz} \citep{Marz2016}.  We obtained
\aatspec spectra with the AAT/2dF instrument between July 2014 and
July 2016.

\item[Magellan/IMACS] IMACS is a multislit spectrograph on
the Magellan Telescope.  We used the IMACS in the f/2 camera mode with
the 300 l/mm grism and central wavelength $6700\mbox{\AA}$, covering a
wavelength range $3900-8000\mbox{\AA}$ at a resolution of
$1.3\mbox{\AA}$ per pixel.  Data were reduced using the COSMOS software \citep{2011PASP..123..288D}.
We obtained \imacsspec spectra with Magellan/IMACS between 2013--2014.

\item[GAMA Survey] We include spectroscopy from the DR2
GAMA survey \citep{2015MNRAS.452.2087L} which provides an additional
1995 spectra within the virial radius of 3 unique Milky Way analog hosts.  We
include sources from the GAMA {\tt SpecObj} file with quality flag
$\text{\tt NQ} \ge 3$.
\end{description}

We have so far obtained 17,344 redshifts for unique objects which did not have spectra in either SDSS or GAMA.  This includes redshifts for 12,682 galaxies and 1,610 stars around 8 Milky Way analogs for which we have nearly complete spectroscopic coverage, and an additional 3,052 spectra around 8 hosts with incomplete coverage.   We note that this includes 285 galaxies brighter than $r<17.7$ which did not have spectra in the SDSS spectroscopic catalog.   The SDSS spectroscopic completeness for bright galaxies is roughly 90\% in most of our fields, but two of our primary hosts lie outside of the SDSS legacy spectroscopic footprint.  After SAGA follow-up spectroscopy, we are 100\% complete for galaxies brighter than $r<17.7$ for our eight primary hosts.   For our low redshift targets, the SAGA redshift errors average between 20-25\kms\ for the instrument set-ups above based on repeat measurements.

We combine the spectra above with the spectroscopic
catalogs from SDSS DR12 for SDSS spectra where the SDSS SpecObj flag
$\text{\tt zWarning} = 0$.  In cases where objects have multiple spectra
from different sources, we use the weighted coadded redshift.  In the
rare case where redshifts disagree, we use the redshift measured with
the larger aperture telescope.   The full spectroscopic catalog will be
publicly available on the SAGA~Survey website$^\text{\ref{footnote:sagawebsite}}$ after publication or on request to the authors.

\begin{figure}[t!]
\centering\includegraphics[width=1.02\columnwidth]{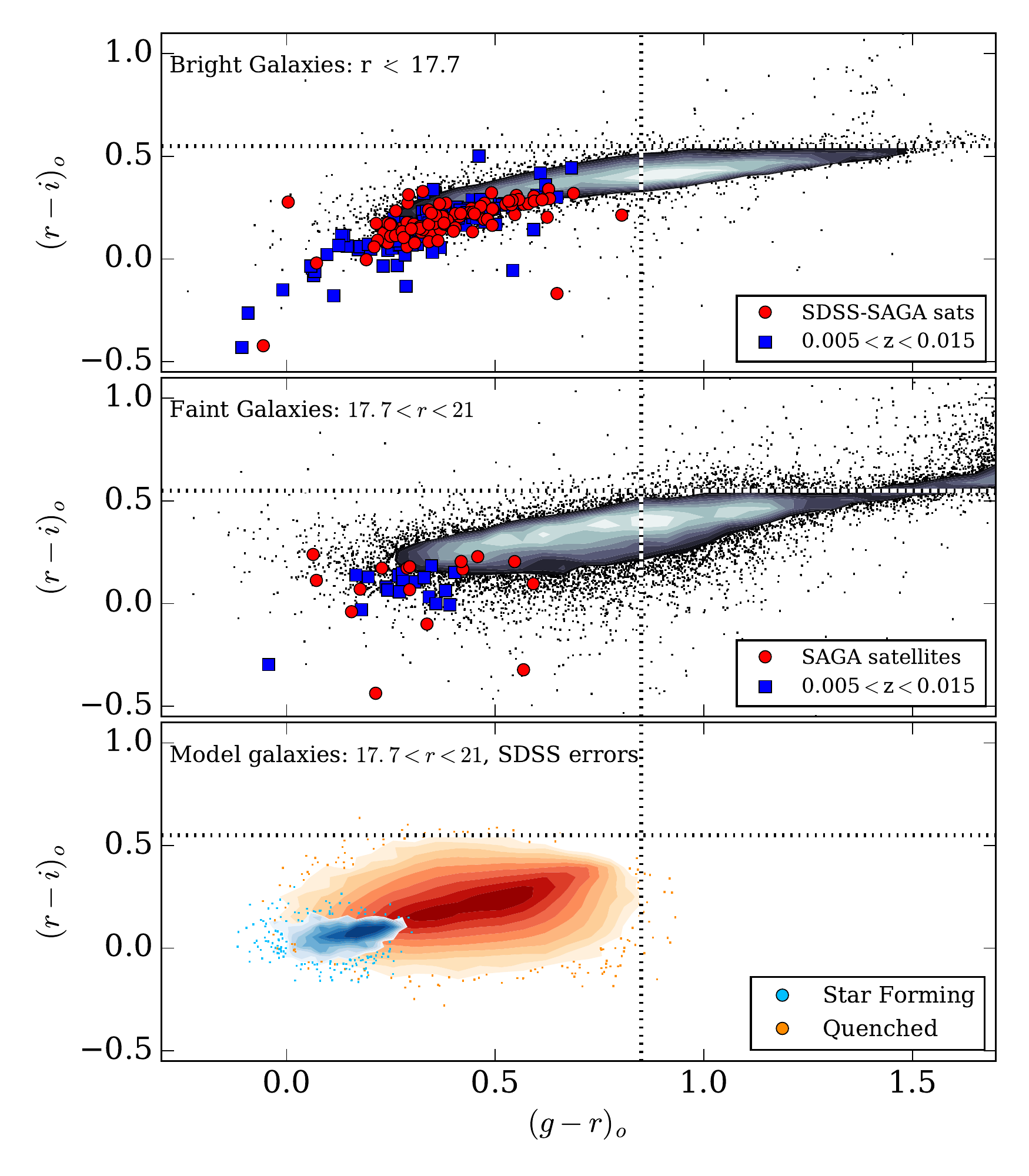}
\caption{Color--color $(g-r)_o$ vs.~$(r-i)_o$ plots including our $gri$ color selection (bottom right quadrant, all panels).  In the top two panels, the contours show the SDSS legacy survey data.   For SAGA objects with spectroscopic follow-up, we differentiate between targets brighter ({\it
    top}) and fainter ({\it middle}) than the SDSS spectroscopic
  magnitude limit of $r_o = 17.7$.  In the upper panels, we plot SAGA satellite galaxies as red circles and a field galaxies over a similar redshift range ($0.005 < z < 0.015$) as blue squares.  
   ({\it Bottom\/})  Semi-analytic model predictions based on \citet{Lu2014}.  Galaxies are simulated between $-12 < M_r < -16$ at $z=0.01$ and plotted for star-forming (blue) and quenched (orange) model galaxies, including realistic SDSS photometric errors.   The vast majority of low redshift
  galaxies, both observed and predicted,  lie within our defined $gri$ color cuts.    \label{fig_colorcolor}}
\end{figure}

\section{Towards an Efficient Method to Select Low Redshift Galaxies}\label{sec_selection}

For Milky Way analogs between $20-40$\,Mpc, there are typically several thousand
galaxies within the projected virial radius (300\,kpc) to our target
depth of $r_o < 20.75$.  This is compared to less than ten satellites
expected, based on the Milky Way, over the same physical region.   Our survey strategy was to obtain complete spectroscopy for 
one Milky Way analog host with no color selection (\autoref{ssec_odyssey}).   We then use these data, in conjunction with ancillary data and theoretical models, to develop a conservative $gri$ color cut which reduces the number of required spectroscopic follow-up targets without risk of removing a low redshift ($z< 0.015$) galaxies (\autoref{ssec_gri}).    We have nearly complete spectroscopy in these $gri$ criteria for 8 Milky Way analog hosts.  We discuss use of these data to develop more efficient satellite selection methods (\autoref{ssec_ugri}) which we plan to apply in observing our full 100 analog sample.

\subsection{Photometric Redshifts Fail at Low Redshift}\label{ssec_photz}

~Nearby faint galaxies (e.g. $20 - 40$\,Mpc, $M_r > -16$) are
difficult to distinguish from the far more numerous background galaxy
population via SDSS photometry alone.  In any given survey there are fewer low redshift galaxies available due to volume effects, limiting the number of training galaxies available.  Furthermore, photometric redshift algorithms
 are typically trained on data which has explicitly color-selected for high redshift galaxies \citep[e.g., the BOSS CMASS sample;][]{CMASS12}.   As a result, widely used
photometric redshifts are unreliable for galaxies with redshifts below $z < 0.015$.    To illustrate this point, in \autoref{fig_photoz} we compare our spectroscopic redshifts to photometric redshifts from the SDSS DR12 \citep{sdss_photoz16}.  While there is rough one-to-one agreement at most redshifts, there is significantly more scatter for both our SAGA satellites (red circles) and a larger sample of `field' galaxies covering a similar, but slightly larger, redshift range ($0.005 < z < 0.015$; blue squares).   Photometric redshifts perform relatively poorly for these two samples, with large fractional uncertainties at all magnitudes (right panel, \autoref{fig_photoz}). While there is \emph{some} correlation in the sense that the photometric redshifts for these samples  cluster towards lower redshifts, there is a well-populated tail with incorrectly high photometric redshifts. As detailed in \autoref{sec_incomplete}, the SAGA~Survey places a premium on high completeness for the satellites, so the existence of this tail prevents our use of photometric redshifts as even a secondary method to increase efficiency.

In order to develop an efficient yet complete candidate selection algorithm for
satellite galaxies, we require an unbiased training set down to our
target magnitude of $r_o < 20.75$.  While spectroscopic surveys exist
to these depths, literature data are either color-selected for high redshift
galaxies \citep[e.g., DEEP2; ][]{DEEP2}, cover too small an area on the sky \citep[e.g., PRIMUS, 9 square degrees][]{PRIMUS11} and/or are too shallow \citep[e.g., GAMA, r < 19.2;][]{2015MNRAS.452.2087L}.  We have chosen to obtain
our own spectroscopic training data, supplemented by the literature where possible.

\begin{figure}[t!]
\includegraphics[width=1.0\columnwidth]{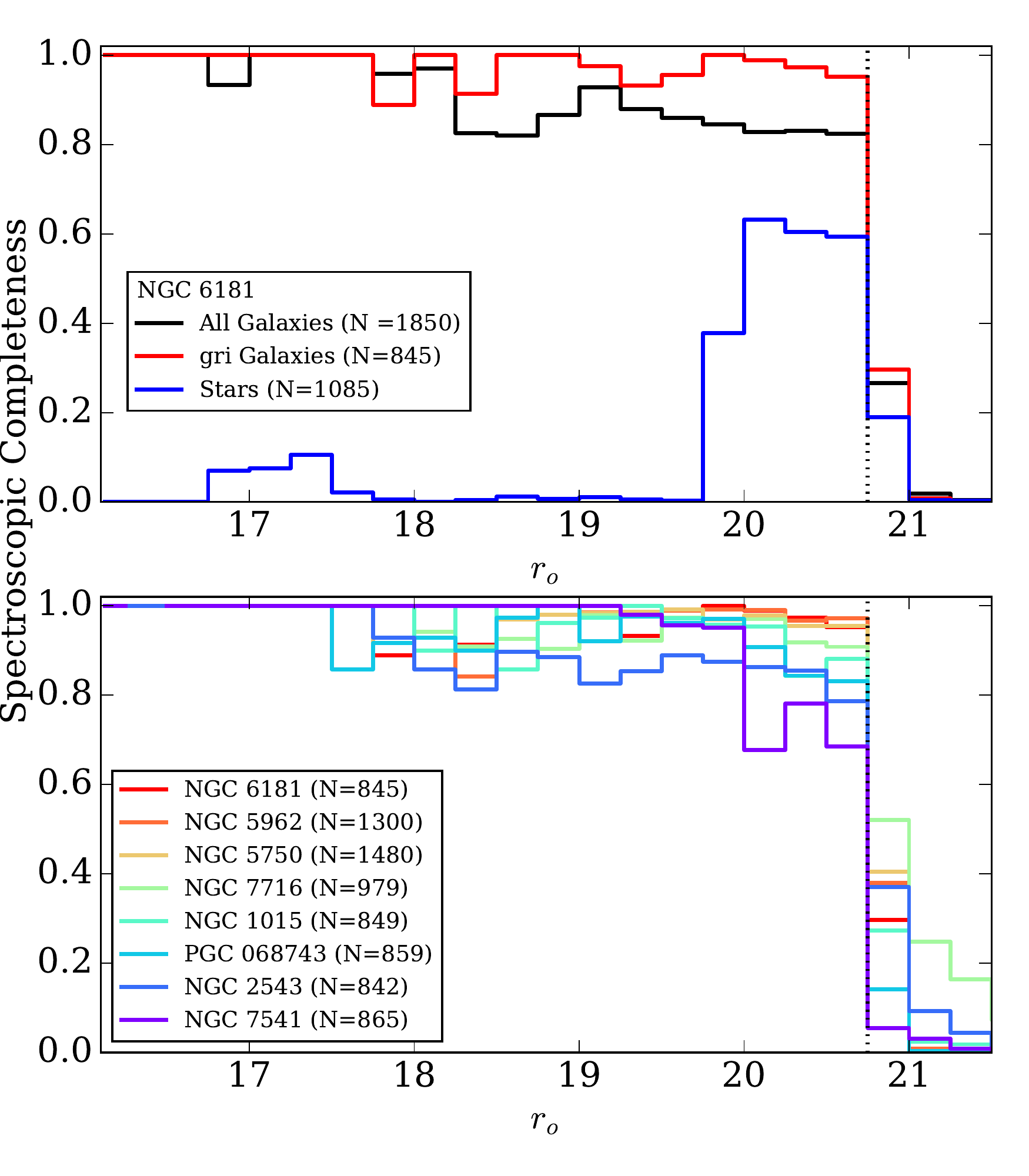}
\caption{{(\it Top\/)} Spectroscopic completeness for targets in
  NGC\,6181 as a function of extinction-corrected $r$-band magnitude.
  The completeness for all galaxies (black), $gri$-selected galaxies (red) and stars (blue) inside of the virial radius down to our
  magnitude limit of $r_o = 20.75$ (dotted vertical line).  {(\it
    Bottom\/)} Spectroscopic completeness for galaxies passing our
  $gri$ color cuts for our \nhost top hosts.  The total number of
  galaxies passing our $gri$ criteria is listed in the legend for each
  host. \label{fig_complete}}
\end{figure}

\subsection{Complete Spectroscopy for NGC 6181}\label{ssec_odyssey}

For the first Milky Way analog in the SAGA~Survey, NGC 6181, we aimed to measure spectroscopic redshifts for galaxies in our survey region without imposing a color criterion.   We obtained spectroscopy for the majority of objects classified by SDSS as galaxies (1580 out of 1850, 85\%) within the virial radius down to $r_o < 20.75$.  A more detailed discussion of data for this individual system will be in a forthcoming paper by Weiner et al.~(in preparation).

We plot spectroscopic completeness for the NGC~6181 sample as a function of $r$-band
magnitude (black line, top panel \autoref{fig_complete}), achieving greater than 80\% completeness in any given magnitude bin.   Plotting these data in $gri$ color-space (and combining the less complete data from
our other hosts) in the top panels of \autoref{fig_colorcolor}, we see that both
the satellites and non-satellites in the same redshift range ($0.005 < z < 0.015$) cluster towards blue $gri$ colors.  This motivated our $gri$ color selection below.  We additionally obtained spectroscopy for 1085 faint stars in NGC 6181, to check for possible missed objects due to star/galaxy separation issues, which are discussed in \autoref{ssec_m32}.

\subsection{The gri Sample}\label{ssec_gri}

We design a $gri$ color cut to safely remove high redshift galaxies and reduce target density, without sacrificing completeness for nearby galaxies $0.005 < z < 0.015$.  Our color selection was designed to include all of the NGC~6181 satellites, field galaxies throughout the SDSS in the same redshift
range and theoretical predictions of $gri$ colors in this regime.    Our criteria are as follows:
\begin{equation}
(g_o - r_o) - 2 \sqrt{\sigma_{g}^2 + \sigma_{r}^2}  < 0.85,
\end{equation}
\begin{equation}
(r_o - i_o) - 2 \sqrt{\sigma_{r}^2 + \sigma_{i}^2}  < 0.55,
\end{equation}
where $\sigma_{g,r,i}$ are the one sigma photometric errors as reported by SDSS. These criteria define the lower left region in each panel of \autoref{fig_colorcolor}.  

All satellites identified around our \nallnoflag SAGA~Survey analog galaxies (red circles, top panels of \autoref{fig_colorcolor}) pass our $gri$ criteria.   This includes 91 satellites brighter than $r_o < 17.7$ observed in SDSS, and 18 faint satellites found in SAGA spectroscopic follow-up.   In addition, there are 205 bright and 35 faint field galaxies in the SAGA survey regions which pass our $gri$ cuts in the same redshift window ($0.005 < z < 0.015$; blue squares, top two panels of \autoref{fig_colorcolor}).  To check that our $gri$ color cuts are not missing faint low redshift galaxies in a larger sample, we search the SDSS CMASS
survey \citep{CMASS12}.  This search includes redshifts for over 1.1 million galaxies between $17.7 < r < 21$.  The CMASS sample targets galaxies which lie in the opposite
quadrant of our $gri$ sample \citep{CMASS12}.   We find that
20 of these galaxies (0.002\%) pass our definition of low redshift.  A
visual inspection of these objects reveals that the majority are
shreds of large galaxies (which we attempt to correct in our own
photometric catalogs, as described in \autoref{sssec_sdssphot}).   The remaining objects are red quasars whose high redshift spectrum has been mis-identified as low redshift.  This extremely small fraction of objects suggests that our $gri$ cuts are nearly complete. 

To further confirm that our $gri$ color cuts are complete at low redshift, we compare to predicted $gri$ colors from semi-analytical models.  We use models based on \citet{Lu2014} which employs flexible parameterizations for the baryonic processes of galaxy formation to encompass a wide range of efficiency for star formation and feedback.  The model is applied to a set of halo merger trees extracted from the Bolshoi simulation \citep{Klypin2011a}; the mass resolution tracks galaxies down to a halo mass of $\sim7 \times 10^9 \msun\,h^{-1}$. The model parameters governing star formation and feedback are tuned using an MCMC optimization to match the stellar mass function of galaxies in the local Universe \citep{Moustakas2013a}.
Therefore, it is guaranteed to produce a global galaxy stellar mass function for the stellar mass range between $10^9$ and $10^{12}\msun$ in the local Universe within the observational uncertainty.  Using this model, we generate galaxies at $z=0.01$ with absolute magnitudes in the range $-16 < M_r < -12$ in the distance range $20-40$\,Mpc with realistic photometric errors from the SDSS.  We plot the resulting distribution of $gri$ colors in the bottom panel of \autoref{fig_colorcolor}, differentiating between star-forming (blue-white) and quenched (red-orange) galaxies.   All of the model galaies, both quenched and star-forming galaxies pass our $gri$ criteria.   This supports the case that our $gri$ cuts are not removing low redshift galaxies from our sample.

\begin{figure}[t!]
\centering\includegraphics[width=\columnwidth,trim={0 0.2in 0.4in 0.5in},clip]{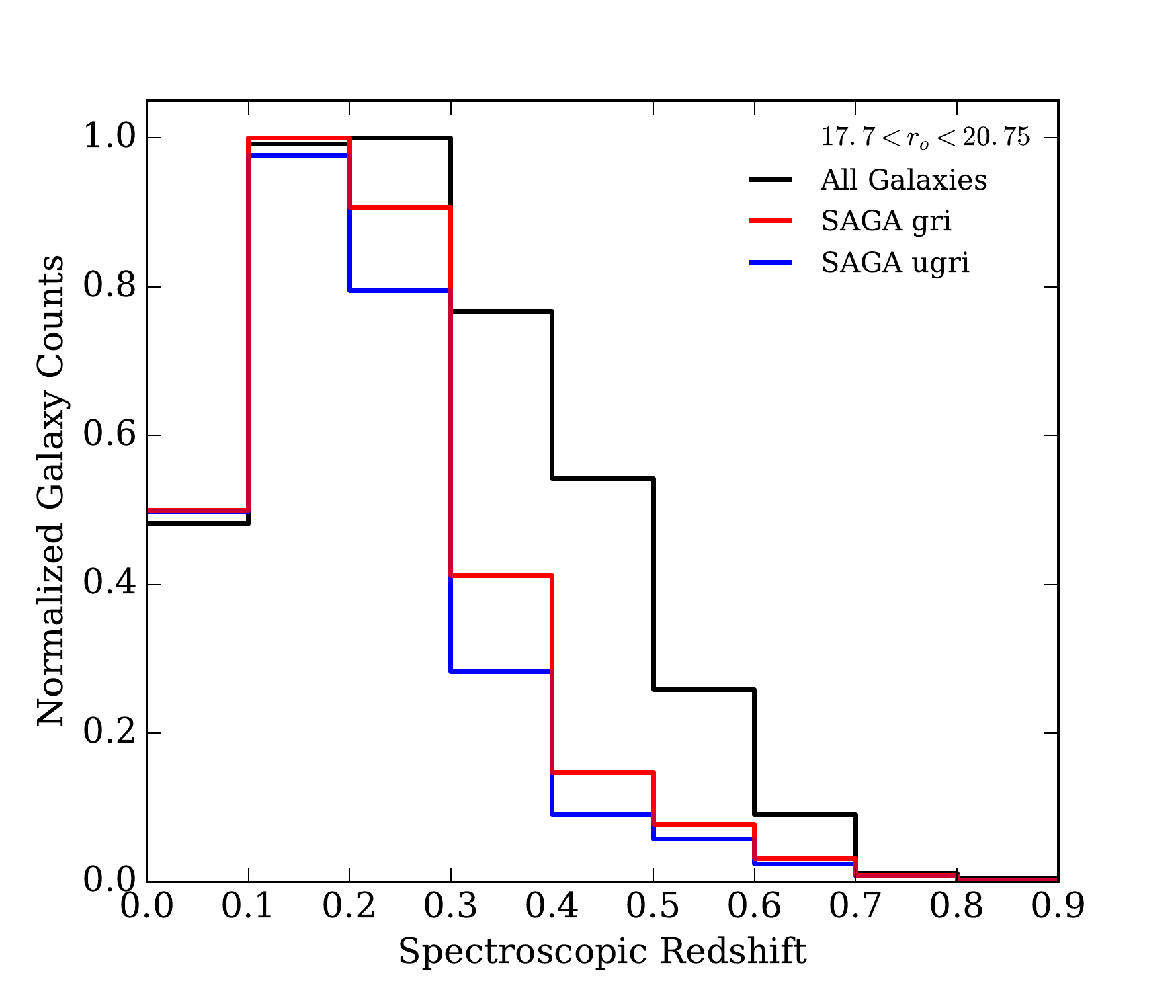}
\caption{Distribution of spectroscopic redshifts for galaxies between
  $17.7 < r_o < 20.75$.  We compare the distribution of all galaxies
  in this magnitude range around NGC~6181 (black) to galaxies passing our $gri$ color
  criteria (red), normalizing the two distributions to peak at unity.
  We show the distribution of our $ugri$ cuts (blue) which further
  reduces the number of higher redshift galaxies without
  affecting completeness at low redshift. 
 \label{fig_redshift}}
\end{figure}

Our $gri$ cuts reduce the number of objects requiring
spectroscopic follow-up by a factor of two or more, from 3000 galaxies per square degree ($r_o<20.75$) to 1250 galaxies per square degree on average.   (See also \autoref{table_hosts}, comparing the number of all galaxies, $N_{\rm tot}$, to the number of $gri$ galaxies, $N_{gri}$, within the virial radius of each host).  The redshift distribution of faint galaxies between
$17.7 < r_o < 20.75$ from the complete NGC~6181 sample is compared to the distribution of our SAGA $gri$
color cuts in \autoref{fig_redshift}.  While the complete distribution
peaks near $z = 0.25$, our $gri$ cut peaks towards lower redshifts at $z=0.15$.  As shown in the
bottom panel of \autoref{fig_complete} and \autoref{table_hosts}, we have achieved higher
than 82\% spectroscopic completeness for these color cuts in \nhost
SAGA hosts, and above 95\% in 4 of these 8 hosts.

\subsection{Improving Efficiency:  ugri and Machine Learning Algorithms}\label{ssec_ugri}

The above $gri$ cuts reduce the total number of candidate
satellites which require spectroscopic follow-up by a factor of two without sacrificing
completeness.   However, over 800 candidate satellite galaxies remain for each host galaxies within the virial radius (\autoref{table_hosts}, Column~11), precluding rapid completion of our 100 analog goal.  We explore whether it is possible to further increase the efficiency of finding satellites by introducing additional observed properties.   

We have explored several additional observed properties.    We find that including an additional cut in $u-r$ color, we are able to further reduce the number of candidate satellites without reducing completeness:
\begin{equation}
(u_o - g_o) + 2\sqrt{\sigma_{u}^2 + \sigma_{g}^2}  > \\ 
1.5 \left((g_o - r_o) - 2 \sqrt{\sigma_{g}^2 + \sigma_{r}^2}\right)
\end{equation}
As shown in the blue curve of \autoref{fig_redshift}, this additional
criterion removes only higher redshift objects in the original $gri$
distribution.  As the SAGA~Survey moves forward, we plan to implement
these $ugri$ cuts in our observing strategy.  While we have also considered cuts on
surface brightness, it is likely that such cuts are harder to replicate in other surveys due to differences in calculating surface brightness and is also a difficult quantity to reproduce in models.  Although SDSS imaging is sufficient to select a complete sample, deeper or better seeing imaging is likely to significantly improve the color cut effectiveness, and would possibly also allow us to effectively use cuts on surface brightness or galaxy size.  

The above selection approach to reducing the number of candidate
satellites requiring follow-up spectroscopy, while conservative,
does not use the imaging data to its fullest extent. 
We are pursuing machine learning
algorithms which can use all SDSS features to efficiently select
targets.  A disadvantage of this approach is that it requires
substantial training data to be effective.  The data presented
in this paper can provide such a training set, and moving forward,
we expect to use such approaches to further improve our selection efficiency.

\begin{figure}[t!]
\centering\includegraphics[width=1.05\columnwidth]{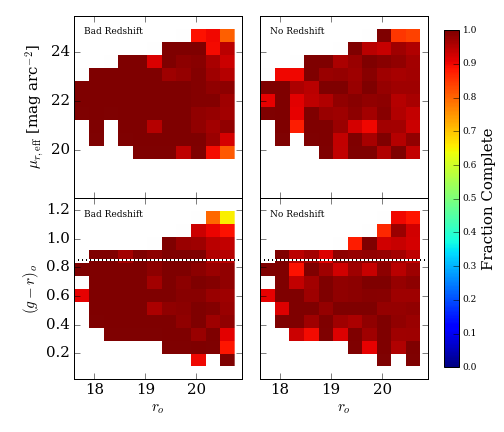}
\caption{We explore possible biases in the galaxies which we did not observe
({\it right}) and galaxies which we observed, but could not measure a redshift ({\it left}) in our 8 SAGA hosts.   We plot the spectroscopic completeness as a function of surface brightness, $\mu_{r,{\rm eff}}$ ({\it top}) or $(g-r)_o$ color ({\it bottom}) as a function of $r$-band magnitude.   We see no strong trends in either of these properties, suggesting that we are not biased against, e.g., low surface brightness or red galaxies due to incomplete spectroscopy. \label{fig_spec_incomplete}}
\end{figure}

\section{Survey Completeness}\label{sec_incomplete}

It is essential that our survey be complete, or have well quantified
incompleteness, down to our stated magnitude limits, since missing even
a single satellite galaxy around a Milky Way host could bias
interpretation.  We showed in \autoref{ssec_photz} that determining whether or not an object is a
satellite requires a spectroscopic redshift.  In
\autoref{ssec_gri}, we motivated spectroscopic follow-up of only galaxies which pass our $gri$ color cuts, and argued that these cuts do not remove low redshift galaxies whether star-forming or quenched.   We present results in \autoref{sec_results} for 8 hosts in which we have obtained nearly complete spectroscopy within the host virial radius for our $gri$ cuts.  Here we explore remaining sources of incompleteness in this sample: galaxies in our $gri$ sample for which we intended to get a redshift, but either did not target or were unable to measure a redshift (\autoref{ssec_incomplete_spec}), and objects which
are not in the SDSS photometric catalogs due to low surface
brightness (\autoref{ssec_incomplete_sb}) or classification as an unresolved object (\autoref{ssec_m32}).  We discuss each of these sources of incompleteness below.

\subsection{Spectroscopic Incompleteness}\label{ssec_incomplete_spec} While spectroscopic completeness within our $gri$ criteria is above 82\% complete for 8 Milky Way analog hosts, and above 90\% complete for 6 of these hosts, we want to verify that the few percent of objects which did not get a redshift are not biased in some way.  In the bottom panel of \autoref{fig_complete}, we plot spectroscopic incompleteness as a function of $r$-band magnitude for our 8  hosts.   While our incompleteness is relatively constant with magnitude for most of our hosts, we note that NGC 7541 is biased in $r$-band magnitude, with incompleteness dropping in the faintest magnitude bins.  The total number of objects without redshifts can be calculated from the last column of Table~1; these objects fall roughly equally into those that we did not observe and those we observed but could not measure a redshift.

A concern at all magnitudes is that we may be preferentially unable to measure redshifts for galaxies with low surface brightness and/or redder galaxies with absorption-line only spectra.  In \autoref{fig_spec_incomplete}, we plot the spectroscopic completeness as a function of photometric properties, differentiating between galaxies that we did not observe (right panels) and those we observed but could not measure a redshift (left panels).  We plot only bins which contain 10 or more galaxies.   In the left panels of \autoref{fig_spec_incomplete}, the fraction of galaxies for which we could not measure a redshift is slightly larger towards fainter magnitudes, as would be expected.   However, there is no strong trend with surface brightness or color:  both high and low surface brightness, and red and blue galaxies had a similar number of redshift failures at the faint end of our survey.  The most incomplete bin is for very faint red galaxies --- these galaxies also have large photometric errors (they are 0.2--0.3\,mag redder than our $g-r$ criteria).  We stress that both our star-forming and quenched galaxies show stellar absorption-line features (Figure~\ref{fig_spec1} to \ref{fig_spec3}), suggesting that we are not preferentially missing quenched galaxies, rather these are objects with large photometric errors that are likely fainter than our magnitude limit.  In the right panels of \autoref{fig_spec_incomplete}, the fraction of objects which we did not target is roughly the same as a function of magnitude, color and surface brightness, supporting our statement that we did not preferentially target galaxies based on luminosity.   These plots imply that we are not missing, e.g., low surface brightness or absorption-line satellites due to incomplete spectroscopy.   Given these distributions, we will use the spectroscopic incompleteness as a function of magnitude to correct our luminosity functions as discussed in \autoref{sec:LF}.

\begin{figure}[t!]
\centering\includegraphics[width=\columnwidth]{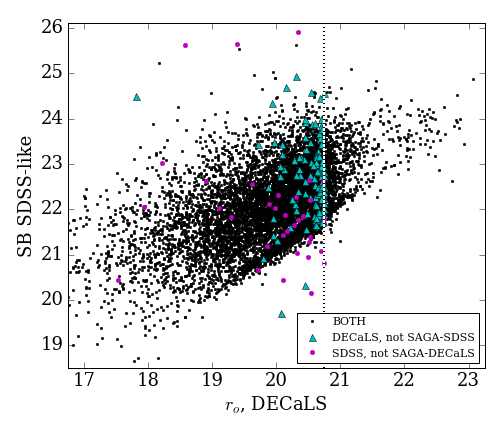}
\caption{ SDSS-like surface brightness (within half-light radius as described in \autoref{ssec_incomplete_sb}) versus $r_o$-band magnitude  using photometry from DECaLS. Black circles are those classified as galaxies in both SDSS and DECaLS, magenta are galaxies in the SDSS with no matches in DECaLS, and cyan are galaxies in DECaLS that have no matches in SDSS (both within $3^{\prime \prime}$). Visual inspection reveals that all of the magenta points are real galaxies that DECaLS failed to detect due to image artifacts, while most of the cyan points are in SDSS but with magnitudes just below the cutoff (see \autoref{ssec_incomplete_sb} for more details).  \label{fig_decals}}
\end{figure}

\subsection{Incompleteness in the SDSS Photometric Catalog}\label{ssec_incomplete_sb}

The SAGA~Survey spectroscopic limit of $r_o < 20.75$ (i.e., for most fields, $r \lesssim 21$) is comfortably
brighter than the SDSS photometric limits.  However, since we are
interested in faint satellite galaxies, it is possible that
the SDSS photometry is missing targets which pass our magnitude limit but are not detected due to low surface brightness.   To check for the presence of low surface brightness galaxies that might be missing from the SDSS, we compare to overlapping areas of the Dark Energy Camera Legacy Survey\footnote{\urlrm{legacysurvey.org}} \citep[DECaLS; ][]{decals2016}.
DECaLS is a wide-field optical imaging survey using the Dark Energy Camera on the Blanco Telescope.  The survey target depth is $r=23.9$, however this region of the imaging data was taken by the Dark Energy Survey \cite{des}, and is therefore deeper.  The catalog is produced using the Tractor inferential source detection algorithm \citep{decals_tractor}.

Only one of the SAGA hosts (NGC\,7716; AnaK) has complete $gr$ imaging within the DECaLS DR3 footprint.  Thus, DECaLS photometry is not yet suitable for actual targeting of SAGA satellites, but  does provide an opportunity to use the deeper (and better seeing) DECaLS data to determine if the SDSS data is missing significant numbers of potential satellites brighter than our detection limit.   We compute a surface brightness within the DECaLS catalog reported effective radius (either de Vaucouleurs or exponential, depending on the best-fit profile), using linear (flux) interpolation along the DECaLS catalog aperture magnitudes.  While not identical to the SDSS surface brightnesses due to the lack of identical information and algorithms in the DECaLS catalog, this provides a reasonably close match for objects detected in both the SDSS and DECaLS.

In \autoref{fig_decals} we plot this DECaLS surface brightness against the DECaLS $r$-band magnitude, corrected for extinction in the same manner as SDSS.  We are interested in all galaxies which pass our SAGA-defined magnitude limit of $r_o < 20.75$ in either the SDSS or DECaLS catalog.   We first match our SAGA-SDSS galaxy sample to the DECaLS catalog within a 
$3^{\prime\prime}$ radius.   We find 8848 matches and plot these as black symbols in \autoref{fig_decals}.  We note that the DECaLS photometry is in general agreement with SDSS, although there is a tail of objects which DECaLS measures as a magnitude or more fainter than SDSS.  These are primarily objects with poor sky subtraction which were not caught by our cleaning algorithm described in \autoref{sssec_sdssphot}.  There are 35 objects in our SDSS-SAGA catalog that do not have matches in DECaLS.   These are all galaxies which DECaLS missed due to bad photometry or proximity to a bright star.   We plot these objects as magenta symbols using the magnitude and surface brightness from SDSS.    Finally, we create SAGA-DECaLS catalog choosing only galaxies with DECaLS-measured $r_o < 20.75$ and match this to our SDSS catalog.   We find that we are missing 120 galaxies, plotted as cyan points in \autoref{fig_decals}.   The majority of these galaxies are in SDSS, but have measured magnitudes fainter than our SAGA magnitude limit.  Visual inspection of these cases suggests it is driven by subtly different choices in the model-fitting approaches and zero-points of the bands.  However, we cannot rule out the possibility that some of these are due to genuine low surface brightness outskirts that are not detected in the SDSS. In one case, the DECaLS data does detect a genuine object that might be a satellite or background galaxy.  However, this is a very small fraction of the sample, and hence there is no evidence for a substantial population of low surface-brightness objects that the SDSS photometry is missing given our photometric selection criteria.

\subsection{Bias Against Compact Galaxies}\label{ssec_m32}

We have chosen to follow-up only objects classified as spatially resolved by the SDSS.  Within our survey volume, analogs of the Milky Way satellites would be resolved and classified as galaxies by the SDSS (\autoref{fig_mwsims}).  However, in removing stars from our target list we would miss any analogs to the compact elliptical galaxies such as the M\,31 satellite galaxy M\,32 (brown points in
\autoref{fig_mwsims}).  At the beginning of our survey, we obtained spectra for 1085 stars around NGC~6181, 920 of which were faint, $19.5 <r_o < 20.75$.  We found no unresolved
galaxies.  We also examined all objects brighter than our survey magnitude limit which were classified as stars in SDSS, but classified as galaxies by the DECaLS photometry described above.  We find 30 such cases, the majority of which are close star pairs misclassified as a single galaxy by DECaLS.  There are two cases (out of over 10,000 galaxy matches) where SDSS has classified an object as a star, but is marginally resolved in the DECaLS imaging (these two objects are among the cyan symbols in \autoref{fig_decals}).  Although these are far from complete samples, it does suggest that compact elliptical M\,32-like galaxies are rare in Milky Way-like environments.  We
therefore focus the rest of our survey on objects classified in the
SDSS photometry as galaxies.

\begin{figure}[t!]
\centering\includegraphics[width=\columnwidth]{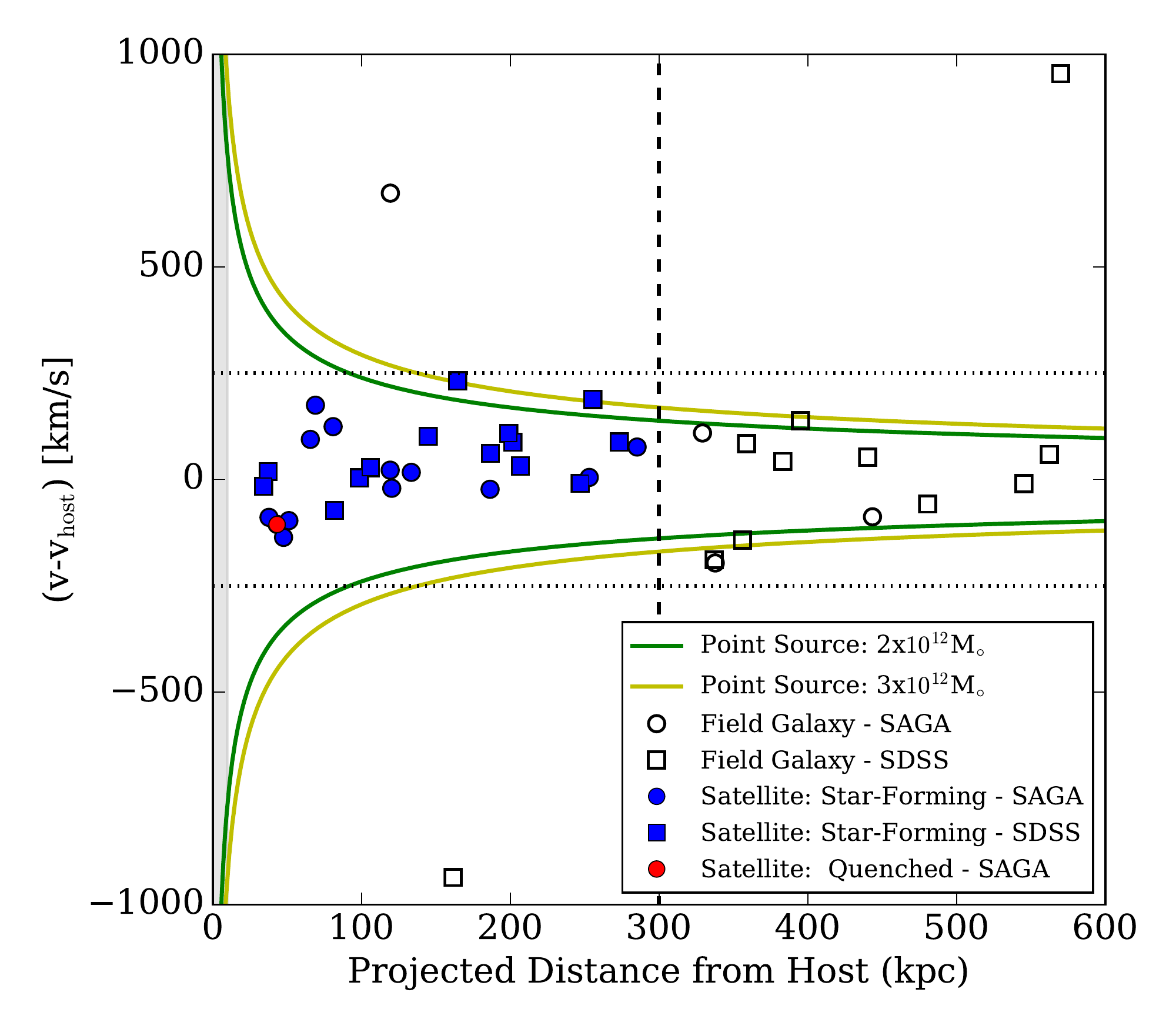}
\caption{Velocity difference versus projected radial distance for 8 complete hosts.  Solid points are defined as satellites within the projected virial radius (dashed line) and $\pm 250$\kms (dotted lines). Open symbols are field galaxies. Solid red/blue indicate satellites which are quenched/star forming.  Square symbols are previously known galaxies, circles are galaxies discovered in this work.  The grey region between 0 to 10\,kpc is excluded due to confusion with the host galaxy.  The solid green (yellow) lines are the escape velocity curve for a $2 \; (3) \times10^{12}\,\msun$ point mass.   \label{fig_vsat}}
\end{figure}

\begin{figure}[t!]
\centering\includegraphics[width=\columnwidth]{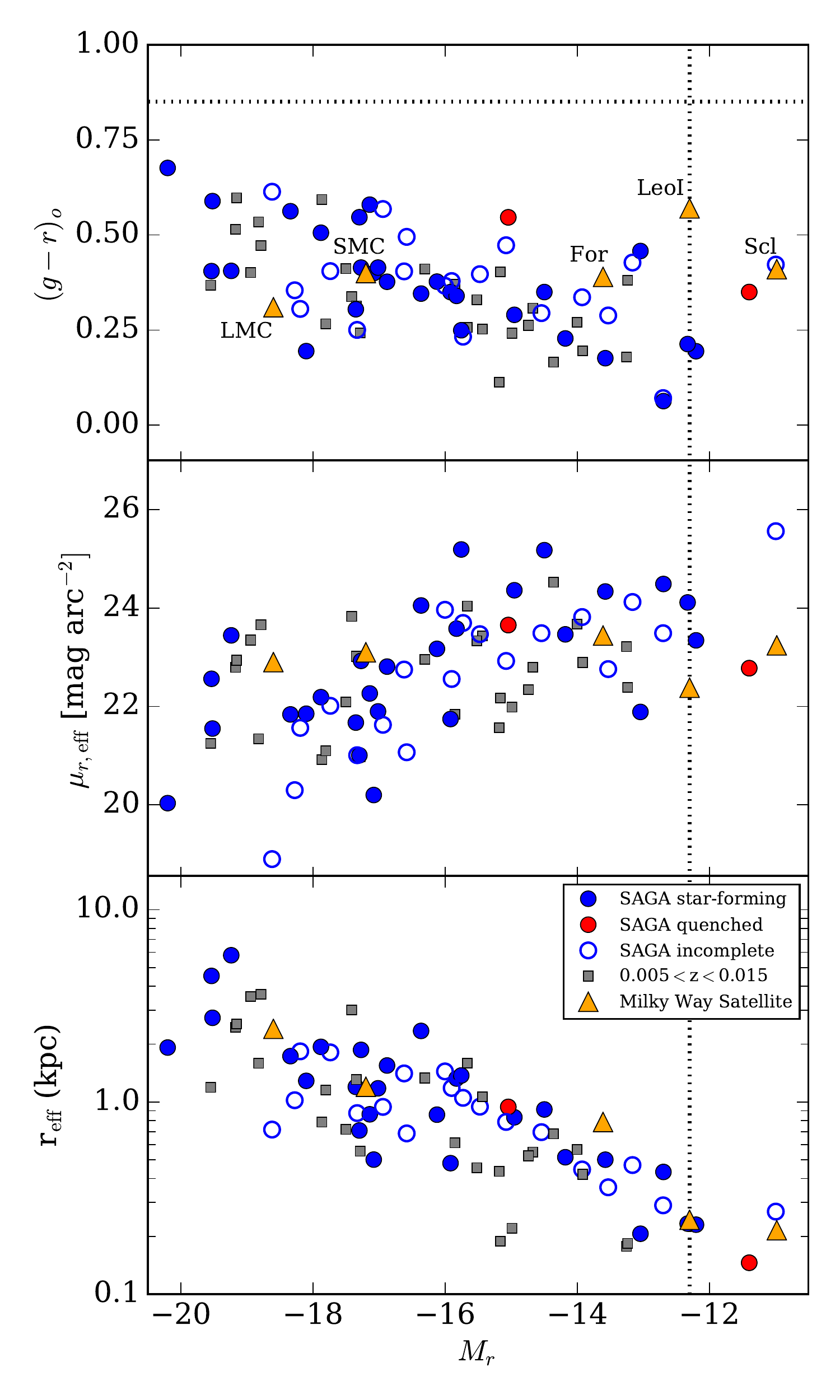}
\caption{Color, effective surface brightness and effective radius (top to bottom panel) plotted against absolute $r$-band magnitude for our SAGA satellites (circles), differentiating between star-forming (blue) and quenched (red) galaxies.  Open circles are satellites in Milky Way analogs in which we do not yet have complete spectroscopic coverage.    We plot field galaxies in the same redshift window (grey squares) and Milky Way satellites (yellow triangles).  The vertical dotted line indicates the magnitude above which we are complete throughout our survey volume.  The horizontal line in the top panel indicates our low redshift color cut at $(g-r)_o < 0.85$.\label{fig_sb}}
\end{figure}

\section{Results}\label{sec_results}

We next present results for \nhost SAGA Milky Way analog host galaxies.  For
these hosts we are at minimum 84\% complete for all candidate satellite galaxies passing
our $gri$ criteria within the virial radius brighter than $r_o < 20.75$
(\autoref{fig_complete}).  We are above 95\% complete for 4 of these host galaxies.   The host properties of all
\nhost systems are shown in \autoref{fig_hosts} and summarized in \autoref{table_hosts}.

\subsection{Satellite Defined}\label{ssec_sat_defined}

We define a ``satellite'' as a galaxy that is within the projected
virial radius (300\,kpc) of the Milky Way analog galaxy and is within
$\pm 250$\kms\ of the host's redshift.  While we considered defining a
satellite based on escape velocity curves (green/yellow curves, \autoref{fig_vsat}), this is slightly more difficult to reproduce in simulations and is no more formally
correct than a straight velocity cut due to the inherent ambiguity of defining ``bound'' satellites in a cosmological context \citep[e.g.,][]{sales07}.  We note the two satellites passing our satellite criteria with velocities consistent with the larger $3\times10^{12}$\,M$_{\odot}$ dark matter halo are associated with two different hosts (see \autoref{table_sats}).  There are no galaxies found between $\pm 250$ to $500$\kms\ of our host's velocity within the virial radius, suggesting both that our hosts have similar halo masses and that these satellites are truly bound to their hosts (e.g., do not have large transverse velocities).  Visual inspection of \autoref{fig_vsat} also appears to show hints of a bias in the velocity distribution with  more satellites at higher velocities relative to the host than lower.  However, this effect is not statistically significant, having a Bayes Factor of $\sim 2$ in \emph{favor} of an equal rather than unequal binomial distribution of velocities. 

Based on our satellite definition, we have discovered 25 satellites.  This includes 14 satellite galaxies meeting our formal criteria around eight complete host systems, and an additional 11 in incompletely surveyed hosts or below our formal magnitude limit. {\it Combined with 13 known satellites, our complete sample includes 27 satellites around \nhost hosts}.  The number of satellites per host ranges from 1 to 9 satellites.  The spatial distribution of satellites around each host is shown in \autoref{fig_images}. 

There are an additional 12 galaxies between 1 to 2 virial radii (seen in \autoref{fig_vsat}), although our completeness varies from 5-35\% per host in this region.  While these galaxies could be bound to the hosts, for the purposes of this paper we classify these as field galaxies.  We expect some contamination in our primary satellite sample due to these nearby field galaxies.   We estimate on average less than 21\% contamination in our satellite numbers due to galaxies between 1 to 2 virial radius, and less than 6\% due to galaxies beyond 2 virial radii.

\begin{figure}[t!]
\centering\includegraphics[width=\columnwidth]{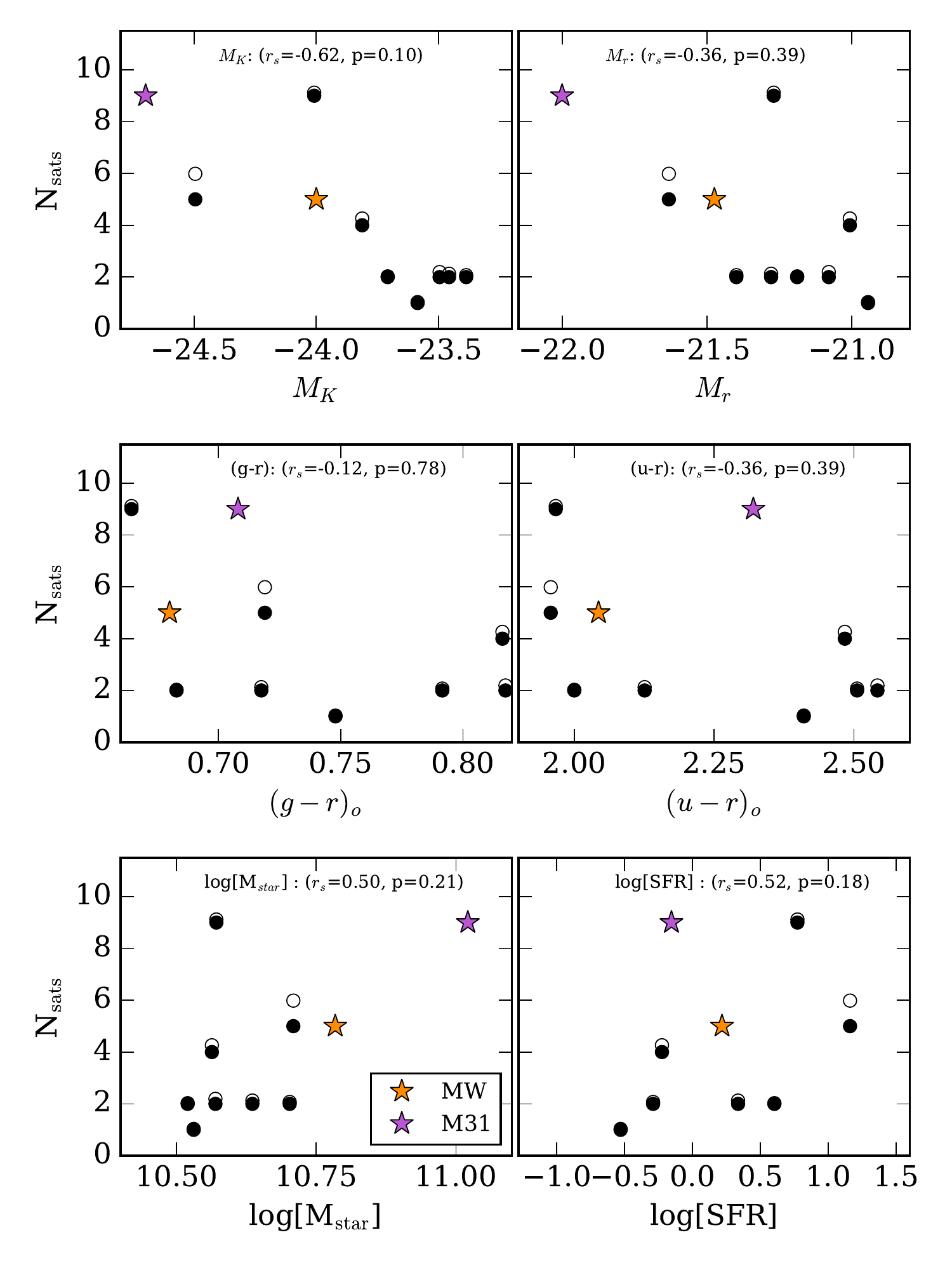}
\caption{The number of satellite galaxies per host brighter than our survey limit of $M_{r,o} < -12.3$ versus various host properties.   Solid black symbols are the observed satellite counts, open symbols are corrected for spectroscopic incompleteness.   The orange star indicates the position of the Milky Way, the purple star is M\,31.   In each panel, we list Spearman's rank correlation coefficient and the corresponding $p$-value as ($r_s$, $p$) for the corrected satellite counts.   While by eye, there appear to significant correlations with $M_K$ and host SFR, these are not significant and will require a larger host sample to confirm.  \label{fig_nsats}}
\end{figure}

\subsection{Satellite Properties:  Star Forming Satellites}

We first compare the properties of our SAGA satellites to those of
the Milky Way.  In \autoref{fig_sb}, we plot color, surface brightness and size of our satellites compared to the Milky Way satellites and our field population at similar redshifts.   For the Magellanic Clouds, we assume sizes, colors and surface brightness from \citet{bothun88} using the photometric transformation of \citet{jester05}.  For the Fornax, Leo\,I, and Sculptor dSphs, we assume properties from Mu\~noz et al.~(in prep.) based on homogeneous wide-field imaging.  We do not include the Sagittarius dSph in \autoref{fig_sb} as it is disrupting and its properties may be compromised.  

The SAGA satellites show the same general trends as the two comparison populations in \autoref{fig_sb}.  Both the sizes and surface brightnesses of the Milky Way satellites are comparable to our SAGA satellites as a function of absolute magnitude.   The main difference between the SAGA satellites and the Milky Way are the colors of the three non-star forming Milky Way satellites (Fornax, Leo\,I and Sculptor).   These are redder than SAGA population, although would still comfortably pass our $gri$ color cuts, and are consistent with colors predicted for quenched model galaxies in \autoref{ssec_gri}.

\begin{figure*}[t!]
\centering\includegraphics[height=0.92\textheight]{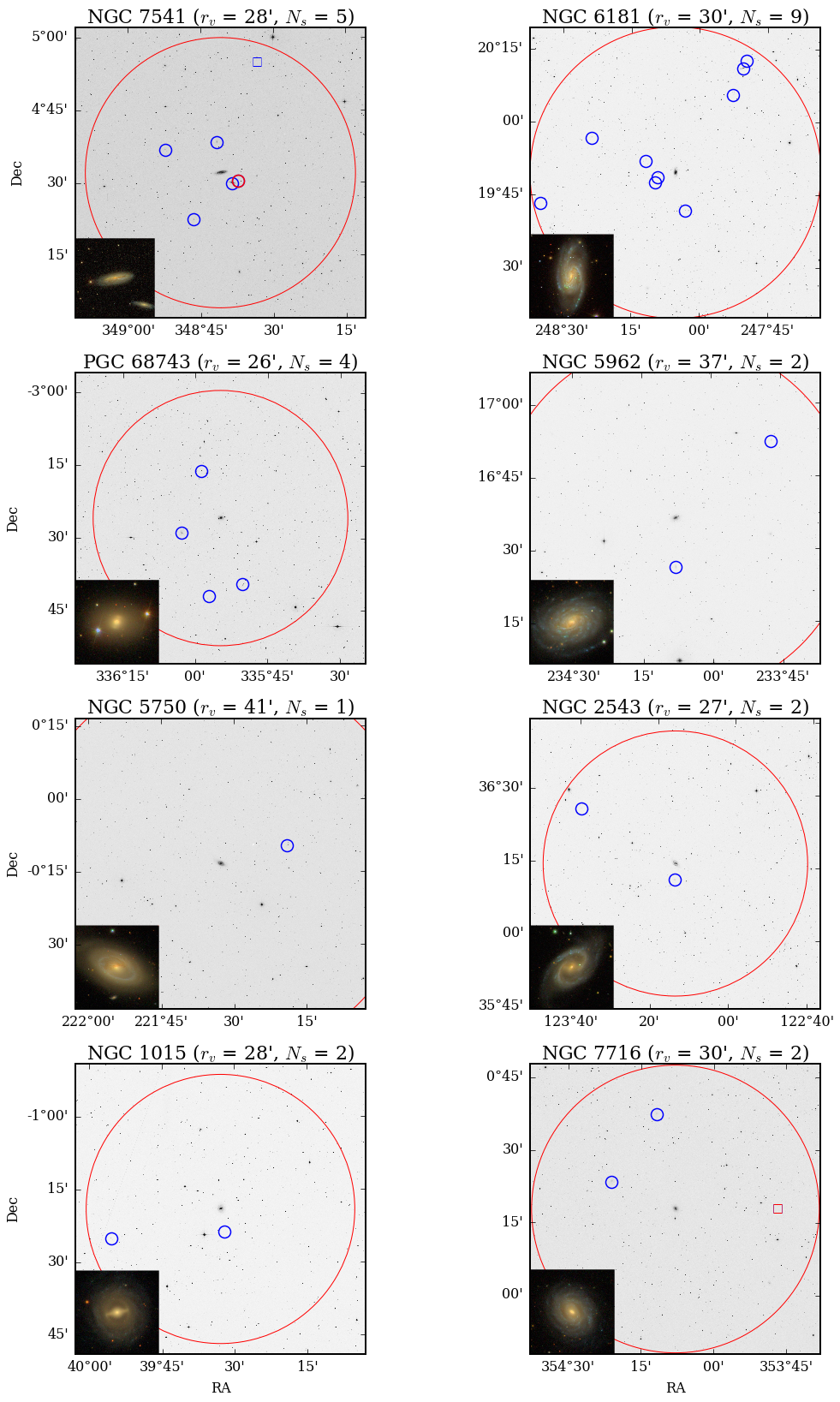}
\caption{Palomar Sky Survey images of the 8 SAGA hosts listed in \autoref{table_hosts} in order of decreasing $M_K$.  Each image is $1^{\circ}$ on a side.  The large red circle indicates the virial radius $r_v$ (300\,kpc at the distance of each host), and is listed in arcminutes in the title of each subplot.  The numbers of SAGA satellites $N_s$ is also listed.  Satellites are plotted as small circles in each panel.  Small blue circles indicate star-forming satellites, small red circles indicate quenched satellites.  Small squares indicate satellites below our completeness magnitude limit.  A SDSS $gri$ image of the host is shown in the bottom left corner of each panel.\label{fig_images}}
\end{figure*}

Perhaps the most surprising result from our survey so far is that the majority of our SAGA satellites are star-forming.   Based on the presence of H$\alpha$ emission in the spectra,  26 out of 27 satellites are star-forming.   The one quenched satellite  ($M_r = -15$) is located in close projection to the brightest satellite ($M_r = -20.1$) of the system, but has a relative velocity of 85\kms (see \autoref{fig_images}), and is thus unlikely to be a satellite of a satellite.   Interestingly, one of the two satellites that lies below our completeness limit is also quenched.

\begin{figure*}[t!]%
\begin{minipage}[t]{\columnwidth}%
\centering\includegraphics[width=\columnwidth, clip, trim={3cm 3.5cm 1cm 4.3cm}]{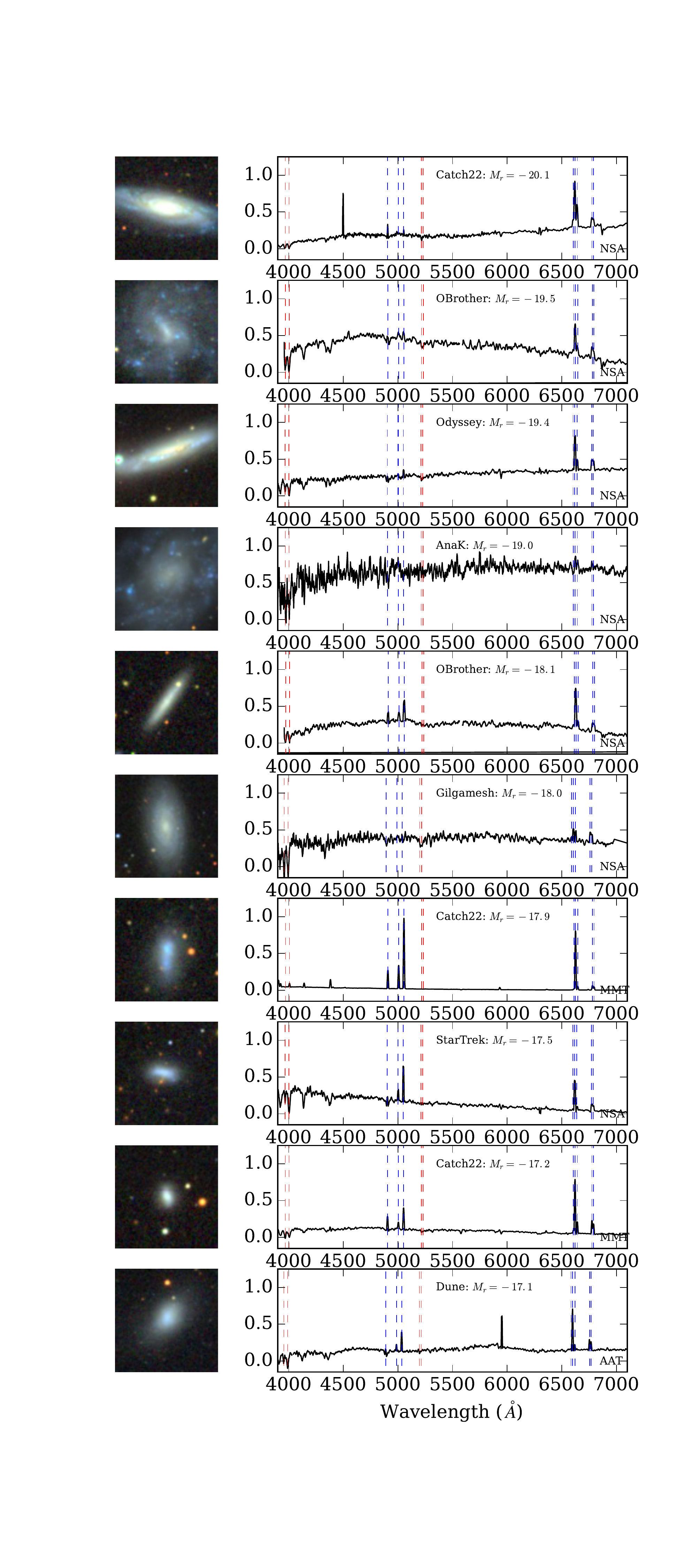}
\caption{SDSS $gri$ composite images generated from the \mbox{DECaLS} Sky Viewer ({\it left}) and associated optical spectra ({\it right}) for the brightest satellites around our complete hosts.  The discovery telescope is indicated in the lower right of each panel.  We indicate common emission (blue) and absorption (red) lines redshifted to the velocity of each galaxy.  We indicate the satellites as `quenched' which show no H$\alpha$ emission.\label{fig_spec1}}
\end{minipage}\hspace*{\columnsep}
\begin{minipage}[t]{\columnwidth}%
\centering\includegraphics[width=0.902\columnwidth, clip, trim={3cm 3.5cm 1cm 4cm}]{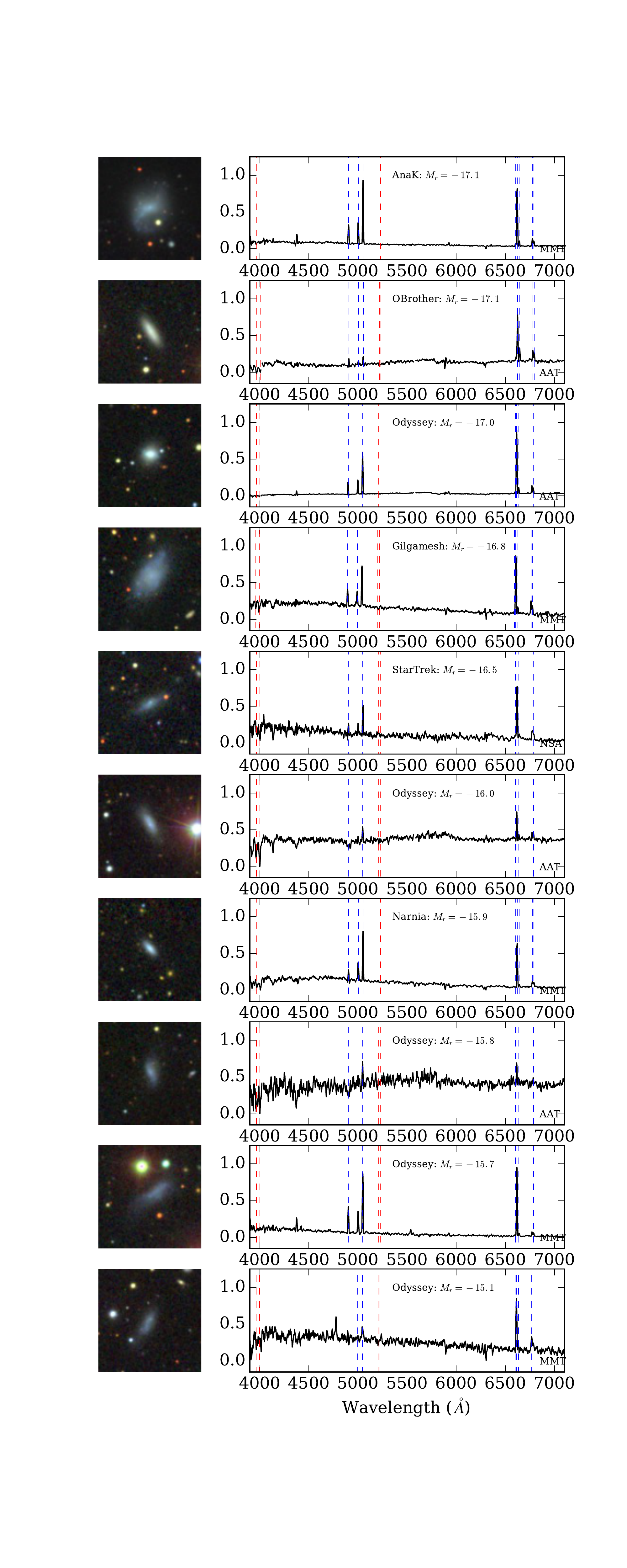}
\caption{Same as \autoref{fig_spec1}, but for fainter SAGA satellites. \label{fig_spec2}}
\end{minipage}%
\end{figure*}

\begin{figure}[t!]
\centering\includegraphics[trim={3cm 3.5cm 1cm 4cm},clip,width=0.92\columnwidth]{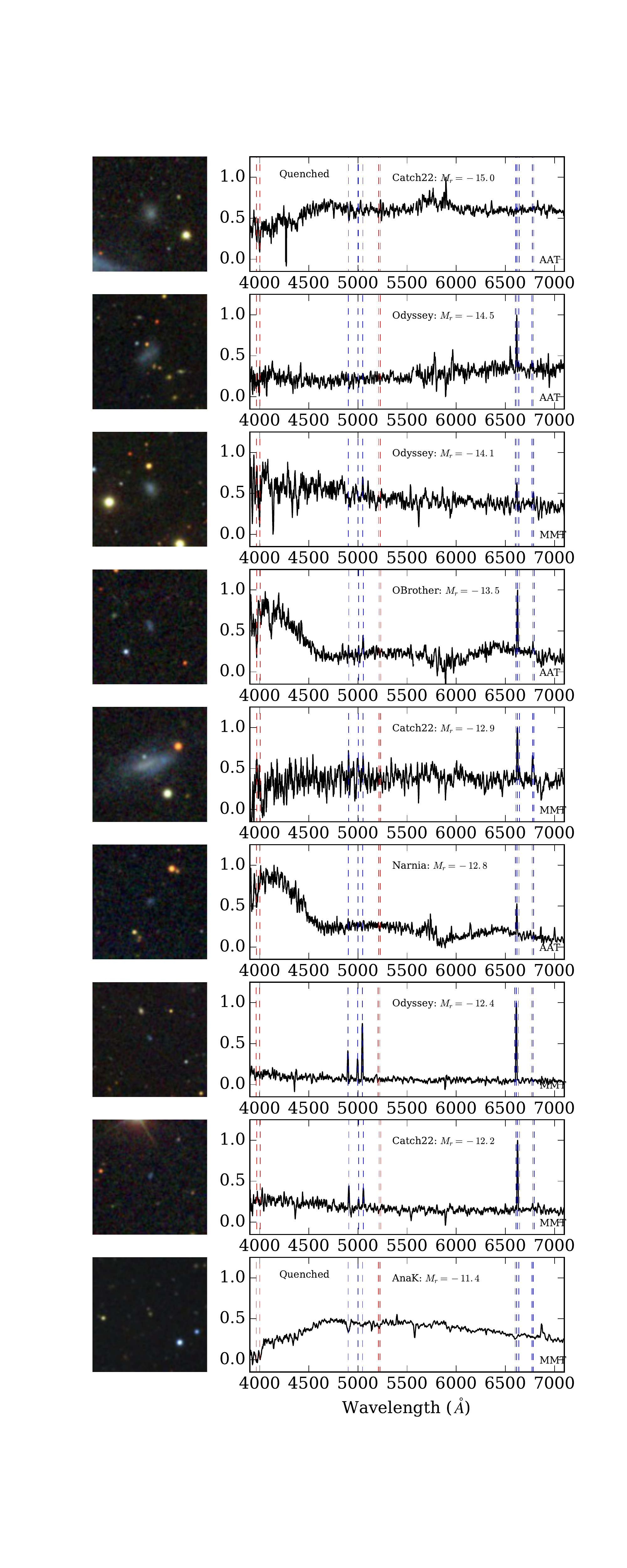}
\caption{Same as \autoref{fig_spec1}, but for the faintest SAGA satellites.  \label{fig_spec3}}
\end{figure}

This large number of star-forming satellites is contrast to both the Milky Way and M\,31 satellite population.   In the Milky Way, only the 2 brightest satellites (the Magellanic Clouds) are forming stars.  In M\,31, there are also only two actively star-forming satellites (M\,33 and IC\,10) in the SAGA luminosity range.  Thus, while 40\% (2 of 5) Milky Way satellites are star-forming and 22\% (2 of 9) of M\,31 satellites are star-forming,
 we find 96\% (26 of 27) of SAGA satellites are star-forming.   One concern might be that we are biased against
spectroscopic identification of quenched satellites, however, as
discussed in \autoref{ssec_incomplete_spec} the targets that we were
unable to measure a redshift for are distributed evenly in color and we detect absorption-line features in our star-forming spectra.  Spectra for all of our satellites are shown in \autoref{fig_spec1} through \autoref{fig_spec3}.   The quenched spectra are indicated and have similarly high signal-to-noise as the rest of our spectroscopic sample.   A similar result was noted by \citet{Spencer2012} for a Milky Way analog at 8\,Mpc.

We further investigate our quenched satellites in \autoref{fig_vsat}. 
We show the radial distribution of satellites, color coding satellites based on the presence
or absence of emission lines in the follow-up spectra.  The quenched satellite in our main sample lies in projection close to its host, while the quenched satellite below our completeness magnitude is close to the virial radius.   These two satellites are around different hosts, both hosts are themselves star-forming.

The physical origin of this result is not clear. We note that our sample of satellites is complete to a given luminosity limit, and not a given stellar mass; in this regime we would expect that star-forming galaxies could be detected to roughly 
$\sim 10^6~\msun$,
while quenched galaxies are likely to only be detected to $\sim 10^7 ~\msun$.  Nevertheless, in the Milky Way, all three galaxies in the range $-16 < M_r < -12.3$ are quenched, compared to only one out of 11 SAGA satellites in this magnitude range.  In our incomplete hosts (Table~1), we similarly find 18 of 18 satellites are star-forming; 7 of these satellite galaxies are in the magnitude range $-16 < M_r < -12.3$.  However, given our small sample size, it is hard to determine whether this result could be unique to a subsample of hosts.  A complete SAGA sample ($\sim$ 100 hosts) will be necessary to make strong statements about the statistical prevalence of quenching in this regime and its dependence on host properties.
Regardless, even the initial sample presented here suggests that satellite quenching may not be as efficient a process as inferred from the Local Group population.

\subsection{Satellite Number and Host Properties}
\label{sec:Nsats}

\begin{figure}[t!]
\centering\includegraphics[width=\columnwidth]{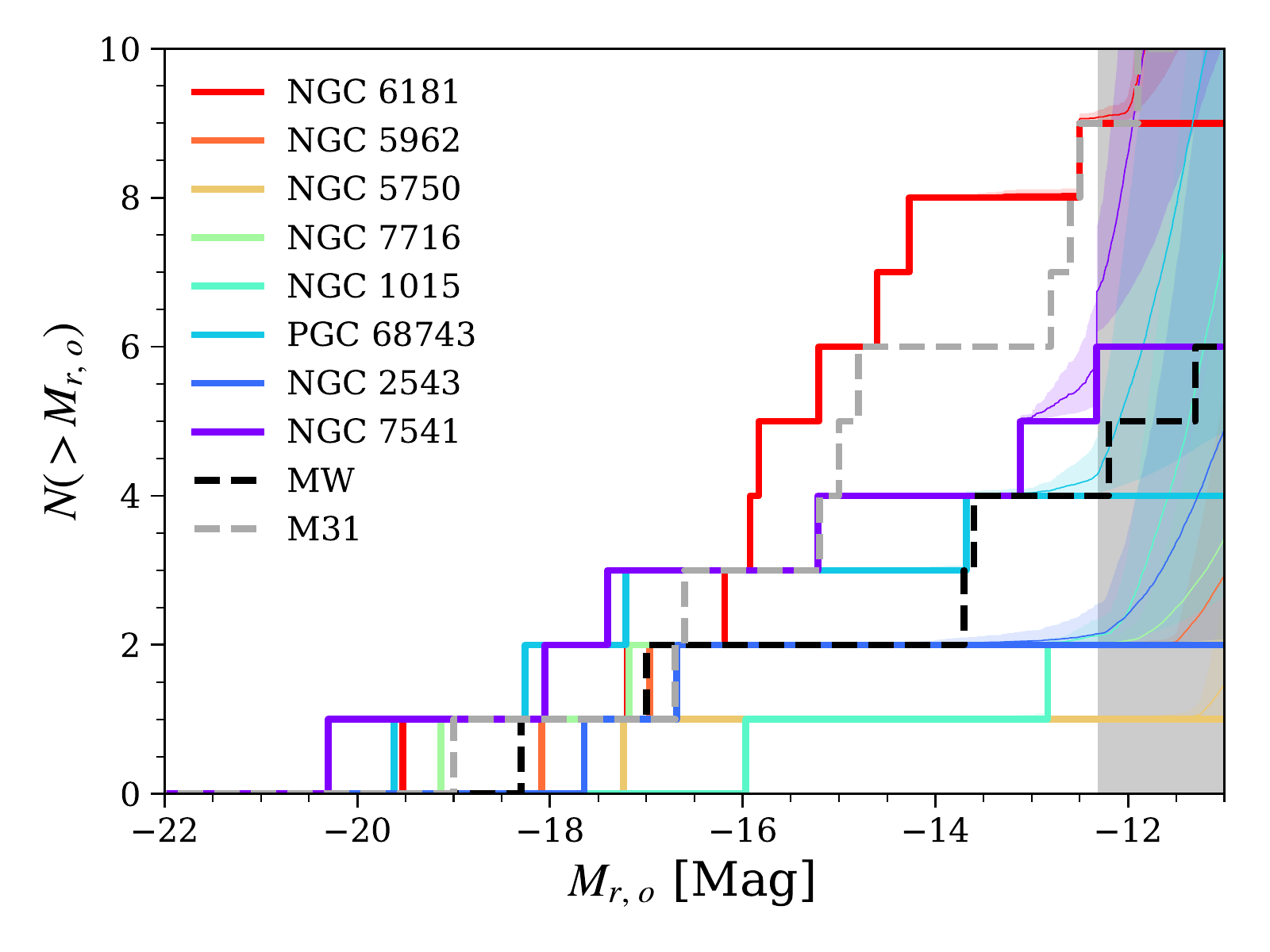}
\caption{The cumulative luminosity function of Milky Way analogs for which we have SAGA spectroscopic coverage (solid lines), as compared to the Milky Way and M31 (dashed lines). 
For MW and M31, the observed luminosities are plotted in $M_V -0.2$ as a proxy for $M_{r,o}$.
Our spectroscopic coverage is incomplete below $M_r > -12.3$ (grey zone). Completeness for brighter satellites varies between hosts (see \autoref{table_hosts}). 
Colored thin lines and bands are the median and the 68\% confidence levels of the completeness-corrected satellite luminosity functions for each corresponding host.
\label{fig_LF}}
\end{figure}

\begin{figure}[htb!]
\centering\includegraphics[width=\columnwidth]{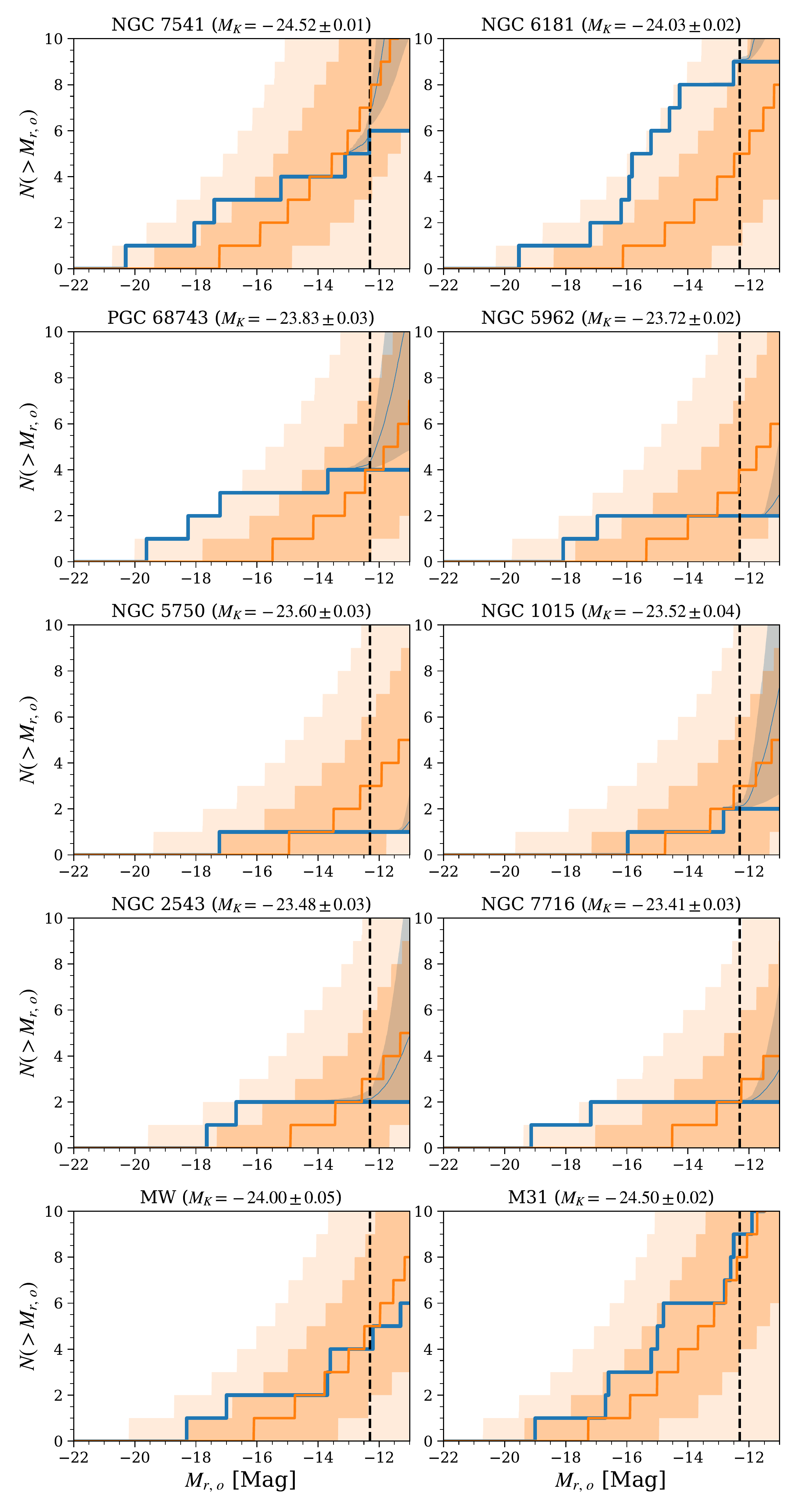}
	\caption{Comparison between the observed (blue) and predicted (orange) satellite luminosity functions for each of the \nhost complete SAGA hosts (sorted by host $M_K$). For comparison the Milky Way and M31 are shown in the bottom two panels. The thick blue lines are the observed satellite luminosity functions.  The thin blue lines and blue bands are, respectively, the median and the 68\% confidence levels of the completeness-corrected satellite luminosity functions. The orange lines and bands are, respectively, the median, 68\%, and 95\% confidence levels of the theoretical prediction from abundance matching. However, there may be other systematic uncertainties that are not captured by these errors.  See \autoref{sec:LF} for details. \label{fig_theory}}
\end{figure}

Before examining the SAGA satellite luminosity functions, we investigate correlations between the number of satellites and host property.   In \autoref{fig_nsats}, we plot the number of satellites per host, $N_{\rm sat}$ (solid black circles), including only satellites brighter than $M_r < -12.3$ for which we are complete throughout our survey volume.  
We plot $N_{\rm sat}$ against various host properties: $M_K$, $M_r$, color, stellar mass and host star-formation rate. The latter two quantities were computed as described in \autoref{ssec_defineMW}. 
We additionally plot the number of satellites corrected for spectroscopic incompleteness as open circles.
To calculate the incompleteness correction, we assume that (1) our sample is complete for $r_o < 17.7$, and (2) for each host, there is a constant probability $P$ that a target ($r_o > 17.7$ and passing our $gri$ cuts) is a satellite.   Given the number of observed targets and number of satellites for each host, we then calculate the posterior distribution of $P$, using a flat prior between 0 and 1, and a likelihood function given by the binomial distribution of success rate $P$. Once we obtained the posterior distribution of $P$ for each host, we assume each of the \emph{unobserved} targets represent $P$ satellites, and construct the completeness-corrected satellite number shown as open symbols in \autoref{fig_nsats}. 
The same correction is also used to plot the completeness-corrected satellite luminosity functions in \autoref{fig_LF} and \autoref{fig_theory}.

While there appear to be visual trends in \autoref{fig_nsats} between satellite number and various host properties, these trends are not significant as measured by the Spearman rank probability listed in the upper right of each panel.  These values were calculated using the completeness-corrected satellite numbers and include the Milky Way.   We calculate the rank correlation, $r_s$, and probability, $p$, of the null hypothesis happening by chance for each panel.   In several cases we see evidence for correlation ($|r_s| \ge 0.5$), but in all cases the probability that the considered correlation happens by chance is 10\% or greater ($p\ge 0.1$), suggesting that more data is needed to determine the strength or existence of any correlation.    As we increase the number of SAGA host galaxies, it will be interesting to see if these trends persist.

The SAGA~Survey will also provide a new perspective on the spatial distribution of satellite galaxies.
The stacked projected radial profile of the satellite galaxies of our \nhost complete hosts resembles that of the Milky Way classical satellites, but is less concentrated. 
In addition, studies have suggested that the satellites in the Milky Way exist in a co-rotating, plane-like structure \citep{Pawlowski2012}, and similar structures are also found in the satellite systems of M31 \citep{Ibata2013}. Efforts have been made to test if this kind of satellite planes is ubiquitous \citep{10.1038/nature13481}, but have not reached a solid conclusion \citep{Phillips2015}.  Using the methods in these two papers, we test the existence of such satellite planes around the SAGA hosts. We find a very weak hint for the existence of such a plane, but at the current stage, these results only indicate that we do not yet have a sufficiently large sample to be statistical meaningful. 

\begin{figure}[htb!]
\centering\includegraphics[width=\columnwidth]{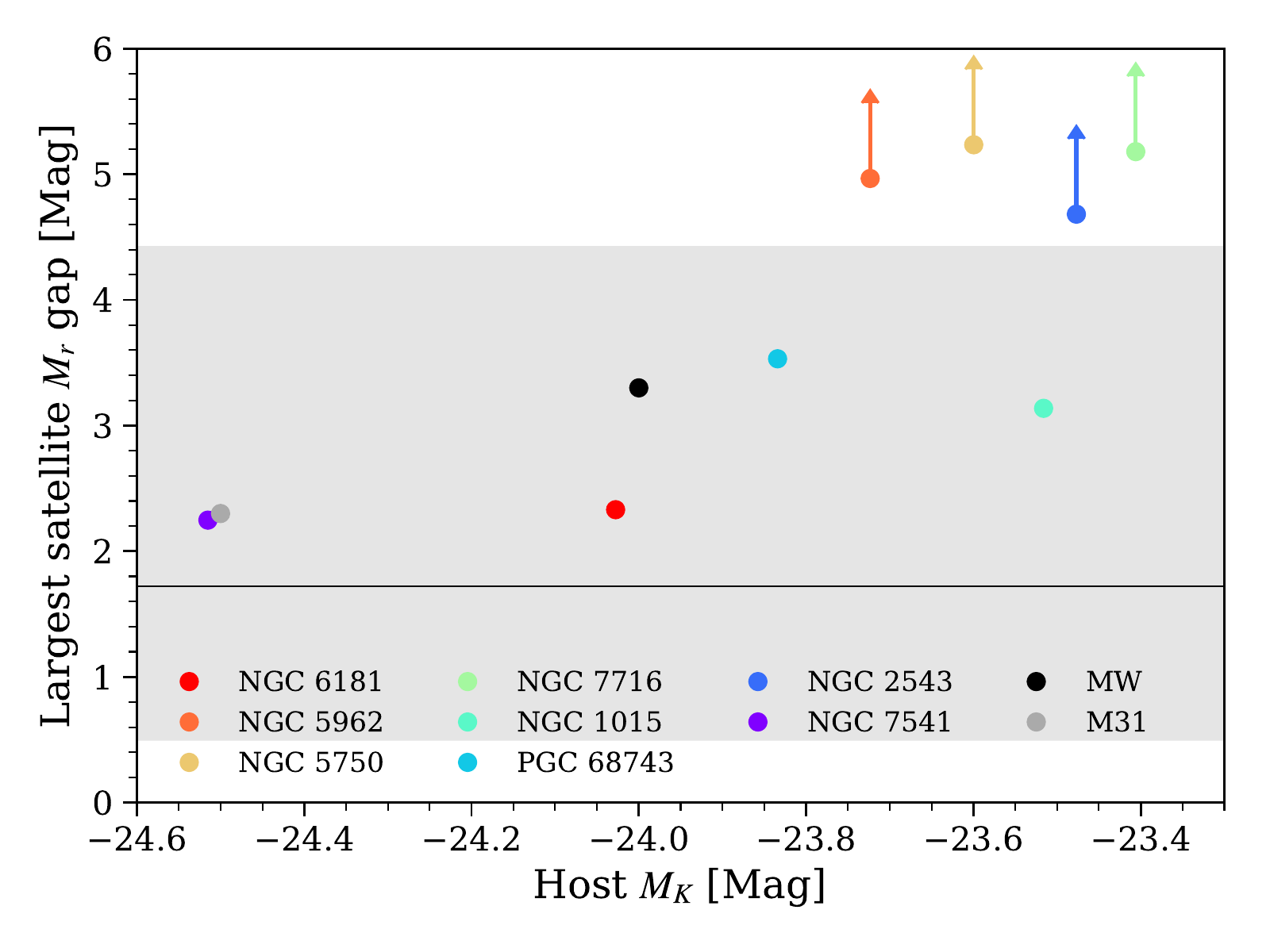}
	\caption{The largest magnitude gap in the satellite luminosity function as a function of host magnitude. The largest magnitude gap does not include the gap between the host galaxy and the brightest satellite. 
    For four of the SAGA hosts, the largest gap extends beyond our spectroscopic coverage limit at $M_r = -12.3$, in which case an up-pointing arrow is added to the point to indicate the largest gap can be in reality even larger. 
    The horizontal gray line and band shows the median and the 95\% ($2\sigma$) confidence levels from the model prediction.
    \label{fig_sat_gap}}
\end{figure}


\subsection{Satellite Luminosity Functions}
\label{sec:LF}

 We present in \autoref{fig_LF} the cumulative satellite luminosity functions for our 8 complete SAGA hosts.
We also plot the completeness-corrected luminosity functions as described in \autoref{sec:Nsats}. 
For comparison we include the satellite luminosity functions of the Milky Way and M\,31.   While the SAGA hosts have similar luminosities as the Milky Way, their satellite systems down to $M_r = -12.3$ are still somewhat different.
In particular, NGC 6181 has a satellite luminosity function as steep as that of M31, which is about twice as massive as the Milky Way. On the other hand, the SAGA hosts have as few as one satellite down to our luminosity threshold.
We also observe that the brightest satellite of each host spans a wide luminosity range. 
The above hints that the satellite luminosity functions of these Milky Way analogs have significant dispersion.

To place our results in a cosmological context, we compare the observed satellite luminosity functions with predictions from a simple $\Lambda$CDM model.  This comparison is shown in \autoref{fig_theory}.  Our model uses the abundance matching technique to (1) find isolated distinct halos that resemble the SAGA hosts based on their $K$-band luminosities (mock hosts), and (2) predict the satellite luminosity functions based on the subhalo mass function (mock satellites). 

We first use the same abundance matching procedure as described in \autoref{ssec_defineMW} to match the 6dF $K$-band luminosity function to the halo peak maximum circular velocity ($\vpeak$) with 0.15 dex of scatter at a fixed $\vpeak$.  For each of the SAGA hosts, we then draw multiple halos that can host galaxies of the corresponding $K$-band luminosity from the halo catalog as our mock hosts.  

We then use the $r$-band luminosity function ($k$-corrected to $z=0$) from the GAMA Survey \citep{loveday2015}, which is measured down to $M_r \sim -12$.
Since the  simulation above does not have sufficient resolution for abundance matching down to $M_r \sim -12$, we extrapolate the matched luminosity--$\vpeak$ relation, where $v_\text{peak}$ is the halo maximal circular velocity at its peak value.
A scatter of 0.2\,dex is applied to the matched luminosity.
We then predict the subhalo $\vpeak$ function based on the host halo properties.
In the SAGA Survey, the satellites are defined to be within 300\,kpc in projection and with a $\Delta V < 250\,\text{km}\,\text{s}^{-1}$.
We apply the same definition of ``satellite'', and fit the normalization of the subhalo $\vpeak$ function assuming a fixed power-law index.
We also assume the normalization depends on host halo mass and concentration only, similar to the treatment presented in \citet{mao2015}.
For each of the SAGA hosts, we run multiple realizations of  subhalo $\vpeak$ functions for each of the mock hosts that corresponds to the SAGA host in consideration, and construct the distribution of the model-predicted satellite luminosity function, plotted in orange in \autoref{fig_theory}.

Comparing the observed satellite luminosity function with the prediction from this simple model provides a new perspective on the ``missing satellite problem'', outside the Local Group.
From \autoref{fig_theory} we notice a few common features. First, although for all hosts presented here the observed luminosity functions are all within 95\% of the ensemble model prediction, the slope of the observed luminosity function is generally flatter then that of the prediction from dark matter simulations.
In many hosts, the model under-predicts the number of bright satellites and over-predicts the number of faint satellites.  Furthermore, despite the weak correlation between the number of satellites and the host galaxy's luminosity, the host-to-host scatter appears to be larger than what the simple model predicts. This may indicate larger scatter in the properties of dwarf satellites for fixed halo properties, scatter in the impact of baryons on the subhalo properties themselves, or some bias in the formation history of the host halos which impacts the population of bright satellites \citep[e.g.,][]{Lu2016}.

The difference in the slope of the luminosity function can be better captured by evaluating the largest magnitude gap in the satellite luminosity function. For example, in the Milky Way, the largest satellite magnitude gap is about 3.3\,mag, sitting between the SMC and the Sagittarius dSph. \autoref{fig_sat_gap} shows the largest satellite magnitude gap of the SAGA hosts and also the MW and M31, as a function of the host luminosity, compared with the model prediction.
Note that in this simple model, we assume a constant slope in the satellite luminosity function, and hence the prediction for the magnitude gap is constant with respect to the host luminosity.
Inspecting \autoref{fig_sat_gap}, we find all hosts have a gap larger than the median of the model prediction, and four of the fainter SAGA hosts have a gap larger than $2\sigma$ from the model prediction. 

These discrepancies between observation and prediction likely indicate some unrealistic assumptions in this simple model.
For example, the extrapolation of the abundance matched luminosity--$\vpeak$ relation may not be valid in this regime.
In particular, the subhalo mass function extracted from dark matter-only simulations may overpredict the actual subhalo abundance, as feedback within dwarf galaxies can change their properties and the neglected baryonic components, such as disks, can further disrupt subhalos \citep{Wetzel2016,Garrison-Kimmel2017}.
Our model is abundance matched to the GAMA global luminosity function, which may differ from the satellite luminosity function of individual host galaxies  \citep[e.g.,][]{Read2017}.
Also, the GAMA global luminosity function itself has systematic uncertainties which we did not include in our model, and these uncertainties can impact the luminosity function, particularly for $M_r > -16$.
Hence, caution should be taken when interpreting the comparison between the observation against this simple model. 
These uncertainties in the model can be parameterized, and then be constrained by these observed satellite luminosity functions. These constraints will provides useful insight on the galaxy--halo connection at this mass scale. 
A larger number of hosts is needed to obtain meaningful constraints, and we plan to continue the survey to reach a sufficient sample.

\section{Summary and Future Survey Plans}

The goal of the SAGA Survey is to determine complete satellite galaxy luminosity functions within the virial radius for 100 Milky Way analogs down to the luminosity of the Leo~I dSph ($M_r = -12.3$).
 We have identified a sample of \nallnosdssflag Milky Way analogs between  $20-40$\,Mpc based on $K$-band luminosity and local environment.   In this work, we present complete satellite luminosity functions for 8 Milky Way analog galaxies.    The main survey results so far are:

\begin{itemize}

\item We measured a total of 17,344 redshifts, including 285 galaxies brighter than $r_o < 17.7$ that were not previously measured by SDSS, 14,506 galaxies with $17.7 < r_o < 20.75$, and an additional 2,553 galaxies dimmer than $r_o>20.75$.

\item Based on this spectroscopy, we have discovered a total of 25 new satellite galaxies.  This includes 14 satellite galaxies meeting our formal criteria around 8 complete host systems, and an additional 9 meeting these criteria in incompletely surveyed hosts or fainter than our survey limit. Combined with 13 satellites already known in the SDSS, there are a total of 27 satellites around the \nhost complete hosts.

\item We have developed simple $gri$ color criteria to more efficiently identify low redshift ($0.005 < z < 0.015$) galaxies.   These color cuts reduce target density by a factor of two without loss of completeness in this redshift range (from an average of 3000 targets per square degree, to 1250 targets per square degree).  With these early results, we now expect to reduce this by an additional 20\% using further cuts in $ugri$ color space; this would result in 1000 targets in a typical host.

\item We have characterized complete satellite luminosity functions for 8 Milky Way analog hosts.  We find a wide distribution in the number of satellites, from 1 to 9, in the luminosity range for which there are five satellites around the Milky Way.  We see no statistically significant correlations between satellite number and host properties, although any correlation would be hard to detect robustly with our small sample size of hosts.

\item  Comparing the observed satellite luminosity functions to a simple $\Lambda$CDM model populated with luminosities using abundance matching, we find larger than predicted scatter in the number of satellites between hosts.   In addition, the slope of the observed luminosity function is generally flatter than that of simple models.

\item  The majority (26 of 27) of SAGA satellite galaxies are actively forming stars. This is significantly different from the Milky Way or M\,31 satellites in a similar magnitude range.

\end{itemize}

The above results suggest that the satellite population of the Milky Way may not be representative of satellite populations in the larger Universe.  Expanding the number of Milky Way analog galaxies with known satellites is required to use these objects as meaningful probes of both cosmology and galaxy formation.

To achieve the SAGA~Survey's goal of satellite luminosity functions around 100 Milky Way analog hosts down to $M_r=-12.3$, we are implementing the $ugri$ color criteria described in \autoref{ssec_ugri},
as well as other more sophisticated methods to increase targeting efficiency such as machine learning
algorithms and deeper, higher spatial resolution imaging.   We expect that our extensive sample of galaxy redshifts with blue $gri$ colors will also have ancillary science applications; these data are available on request or the SAGA~Survey website$^\text{\ref{footnote:sagawebsite}}$.  Measuring the internal velocities of our satellites will be essential for enhancing the interpretation of the data above \citep[e.g.,][]{Jiang2015,Guo2015}, and we are obtaining resolved optical rotation curves and single-dish HI gas measurements for all of our SAGA satellites.   The goal of these efforts is to provide the framework necessary to distinguish between the multiple proposed solutions to small-scale problems in $\Lambda$CDM, and provide an improved understanding of the Milky Way itself in a cosmological context.

\acknowledgments 

We thank Carlos Cunha, Claire Dickey, Yashar Hezaveh, Vikas \mbox{Bhetanabhotla}, Emily Sandford, Jeremy Tinker, and Andrew Wetzel for helpful discussions.  We thank Michael Blanton for his work on the NASA-Sloan Atlas which was critical to the start of this project.   RHW thanks her Physics 16 students for their contributions to visual classification at an early stage of the project.
This work was supported by NSF collaborative grant AST-1517148 awarded to MG and RHW.
MG thanks the John S.~Guggenheim Foundation for generous support.
YYM is supported by the Samuel P.~Langley PITT PACC Postdoctoral Fellowship, 
and was supported by the Weiland Family Stanford Graduate Fellowship.
Support for EJT was provided by a Giacconi Fellowship, and by NASA through Hubble
Fellowship grants \#51316.01 awarded by the Space Telescope Science
Institute, which is operated by the Association of Universities for
Research in Astronomy, Inc., for NASA, under contract NAS 5-26555.
Part of this work uses the computational resources at the SLAC National Accelerator Laboratory, a U.S.\ Department of Energy Office; RHW, YYM, and YL thank the SLAC computational team for their consistent support.   
Observations reported here were obtained in part at the MMT Observatory, a joint facility of the University of Arizona and the Smithsonian Institution.
GAMA is a joint European--Australasian project based around a
spectroscopic campaign using the Anglo-Australian Telescope. The GAMA
input catalog is based on data taken from the Sloan Digital Sky
Survey and the UKIRT Infrared Deep Sky Survey.  GAMA is funded by the
STFC (UK), the ARC (Australia), the AAO, and the participating
institutions.

\software{This research made use of many community-developed or community-maintained software packages, including (in alphabetical order):
Astropy \citep{astropy},
Healpy (\urlrm{healpy.readthedocs.io}),
IPython \citep{ipython},
Jupyter (\urlrm{jupyter.org}),
Matplotlib \citep{matplotlib},
NumPy \citep{numpy},
Pandas \citep{pandas},
\mbox{scikit-learn} \citep{scikit-learn},
and SciPy \citep{scipy}.
This research has also made use of NASA's Astrophysics Data System.
}

\facility{SDSS, MMT (Hectospec), AAT (2dF), Magellan:Baade (IMACS)}

\bibliographystyle{yahapj}
\bibliography{saga_bib,software}

\begin{deluxetable*}{l c c C C C C C c c c c l}
\tablewidth{\textwidth}
\tabletypesize{\footnotesize}
\tablecaption{SAGA Milky Way Analog Hosts\label{table_hosts}}
\tablehead{\colhead{(1)} & 
	   \colhead{(2)} & 
       \colhead{(3)} & 
       \colhead{(4)} & 
       \colhead{(5)} & 
        \colhead{(6)} & 
	    \colhead{(7)}& 
        \colhead{(8)}& 
        \colhead{(9)}& 
	    \colhead{(10)} & 
         \colhead{(11)}&
         \colhead{(12)}&
         \colhead{(13)}\\
	    \colhead{SAGA} & 
        \colhead{NGC}&
	   \colhead{NSA} & 
       \colhead{RA} & 
       \colhead{Dec} & 
       \colhead{Dist} & 
	    \colhead{$M_{r,o}$}& 
        \colhead{$M_{K,o}$}& 
        \colhead{$M_{\rm star}$}&
        \colhead{$M_{\rm vir}$}&
        \colhead{$N_\text{s}$} & 
        \colhead{$N_\text{tot}$}& 
	    \colhead{$N_{gri}$} \\
        \colhead{Name} & 
	   \colhead{Name} & 
       \colhead{Name} & 
       \colhead{(deg)} & 
       \colhead{(deg)} & 
       \colhead{(Mpc)} & 
        \colhead{} & 
	    \colhead{}& 
        \colhead{(log$_{10}\msun$)}& 
        \colhead{(log$_{10}\msun$)}& 
        \colhead{}& 
        \colhead{$r_o<20.75$}& 
	    \colhead{$r_o<20.75$}} 
\startdata
\multicolumn{13}{c}{Complete Hosts}\\
Gilgamesh   & NGC 5962      & 166313 & 234.132   & 16.6078  &  28.0   &  -21.2 & -23.7 & 10.52 &   12.13 &      2 &   2995 & 98\% (1271/1300) \\
 Odyssey     & NGC 6181      & 147100 & 248.087   & 19.8264  &  34.3 &  -21.3 & -24.0   & 10.57 &   12.27 &      9 &   1850 & 97\% (819/845)   \\
 Dune        & NGC 5750      & 165536 & 221.546   & -0.22294 &  25.4 &  -20.9 & -23.6 & 10.53 &   12.08 &      1 &   3557 & 97\% (1433/1480) \\
 AnaK        & NGC 7716      &  61945 & 354.131   &  0.29728 &  34.8 &  -21.4 & -23.4 & 10.70  &   12.01 &      2\tablenotemark{1} &   2356 & 94\% (917/979)   \\
 Narnia      & NGC 1015      & 132339 &  39.5482  & -1.31876 &  37.2 &  -21.1 & -23.5 & 10.57 &   12.05 &      2 &   1976 & 92\% (778/849)   \\
 OBrother    & PGC 68743 & 149781 & 335.913   & -3.43167 &  39.2 &  -21.0   & -23.8 & 10.56 &   12.18 &      4 &   1740 & 90\% (770/859)   \\
 StarTrek    & NGC 2543      &  33446 & 123.241   & 36.2546  &  37.7 &  -21.3 & -23.5 & 10.64 &   12.03 &      2 &   1719 & 85\% (716/842)   \\
 Catch22     & NGC 7541      & 150887 & 348.683   &  4.53406 &  37.0   &  -21.6 & -24.5 & 10.71 &   12.55 &      5\tablenotemark{2} &   2198 & 82\% (706/865)   \\
 \hline
 \multicolumn{13}{c}{Incomplete Hosts}\\
ScoobyDoo   & NGC 4158      & 161174 & 182.792   & 20.1757  &  36.3 &  -20.6 & -23   & 10.31 &   11.89 &      4 &   1471 & 47\% (353/758)   \\
 MobyDick    & NGC 3067      &  85746 & 149.588   & 32.3699  &  25.1 &  -20.2 & -23.1 & 10.19 &   11.90  &      0\tablenotemark{3} &   3635 & 38\% (604/1600)  \\
 Othello     & NGC 5792      & 145729 & 224.594   & -1.09102 &  28.4 &  -21.1 & -24.6 & 10.61 &   12.59 &      2 &   3002 & 26\% (371/1433)  \\
 Alice       & NGC 4030      & 140594 & 180.098   & -1.10008 &  23.2 &  -21.5 & -24.5 & 10.62 &   12.55 &      2 &   5628 & 25\% (657/2681)  \\
 Bandamanna  & NGC 7818      & 126115 &   0.99558 & 20.7524  &  32.5 &  -20.8 & -24.1 & 10.42 &   12.34 &      1 &   2019 & 24\% (230/948)   \\
 Sopranos    & NGC 4045      &  13927 & 180.676   &  1.9768  &  29.5 &  -21.1 & -23.6 & 10.61 &   12.09 &      0 &   3492 & 17\% (314/1888)  \\
 Oz          & NGC 3277      & 137625 & 158.231   & 28.5118  &  24.4 &  -20.8 & -23.0   & 10.44 &   11.88 &      5 &   3801 & 08\% (142/1694)  \\
 HarryPotter & PGC 4948 & 129237 &  20.449   & 17.5922  &  38.7 &  -20.0   & -23.5 & 10.18 &   12.06 &      4 &   1526 & 06\% (53/832)    \\
\hline 
 Milky Way &  - & - & 266.25   & -29.008  &  0.0 &   -21.5   & -24.0 &  10.78 & 12.20 &  5 &   - & -      \\
 M31       &  - & - & 20.449   & 17.5922  &   0.8 &   -22.0   & -24.7 &  11.01 & 12.54 &  9 &   - & -       \\
\enddata
\tablenotetext{1}{A third satellite was discovered in this system at $M_r = -11.3$, but is below the survey  completeness magnitude.}
\tablenotetext{2}{A sixth satellite was discovered in this system at $M_r = -12.2$, but is below the survey completeness magnitude.} 
\tablenotetext{3}{A satellite was discovered in this system at $M_r = -11.3$, but is below the survey completeness magnitude.}
\tablecomments{Milky Way analogs ordered by spectroscopic completeness.   Column (1) lists the SAGA name given to each galaxy for ease of reference.  Columns (2) - (7) are host properties taken from the NASA-Sloan Atlas \citep{NSA}.  Columns (8) is from the 2MASS Extended Source Catalog \citep{2000AJ....119.2498J}.  Column (9) is stellar mass as computed in the NASA-Sloan Atlas; column (10) is host virial mass computed based on the $M_K$ luminosity and abundance matching.  Column (11) is the number of satellites down to the SAGA flux limit of $M_r < -12.3$.  Column (12) is the total number of galaxies within the projected virial radius.   Column (13) is the percentage and number of objects for which we have spectroscopically measured redshifts for sources passing our $gri$ color criteria.  For comparison, properties of the Milky Way and M\,31 are listed in the final two rows.}
\end{deluxetable*}

\begin{deluxetable*}{l c C C C C C c c c R r}
\tablewidth{\textwidth}
\tabletypesize{\footnotesize}
\tablecaption{SAGA Satellite Properties \label{table_sats}}
\tablehead{\colhead{(1)} &    
			\colhead{(2)} & 
            \colhead{(3)} &
            \colhead{(4)} &
            \colhead{(5)} &
            \colhead{(6)} &
            \colhead{(7)} &
            \colhead{(8)} &
            \colhead{(9)} &
            \colhead{(10)} &
	     	\colhead{(11)}&
           	\colhead{(12)}\\
		    \colhead{Host Name} &    
			\colhead{SDSS OBJID} & 
            \colhead{RA} &
            \colhead{Dec} &
            \colhead{$r_{o}$} &
            \colhead{$M_r$} &
            \colhead{$(g-r)_o$} &
            \colhead{$\mu_r$} &
            \colhead{$r_{\rm proj}$} &
            \colhead{H$_{\alpha}$} &
	     	\colhead{$v - v_{\rm host}$}&
            \colhead{Tel}\\
			\colhead{} &    
			\colhead{} & 
            \colhead{(deg)} &
            \colhead{(deg)} &
            \colhead{} &
            \colhead{} &
            \colhead{} &
            \colhead{(mag arc$^{-2}$)} &
            \colhead{(kpc)} &
            \colhead{} &
	     	\colhead{(km s$^{-1}$)}
}
\startdata
 \multicolumn{12}{c}{Complete Hosts}\\
NGC 5962 &      1237665565541728490 &      234.1329262 &       16.440455 &     14.3 &    -18.0 &     0.51 &     22.1 &      81 & Y &     -71.9 & SDSS \\
NGC 5962 &      1237665566078402826 &      233.7870375 &       16.870438 &     15.4 &    -16.8 &     0.38 &     22.5 &     206 & Y &      32.5 & SDSS \\\hline
NGC 6181 &      1237662224092299404 &      247.8400299 &       20.184076 &     13.3 &    -19.4 &     0.59 &     21.8 &     255 & Y &     189.3 & SDSS \\
NGC 6181 &      1237662662147571761 &      248.3932257 &       19.946140 &     15.6 &    -17.0 &     0.40 &     20.1 &     186 & Y &      62.3 & SDSS \\
NGC 6181 &      1237662698115432544 &      248.0513396 &       19.695740 &     16.6 &    -16.0 &     0.38 &     22.4 &      80 & Y &     125.3 & AAT \\
NGC 6181 &      1237662662147310256 &      247.8258922 &       20.210879 &     16.9 &    -15.8 &     0.34 &     23.1 &     273 & Y &      89.3 & AAT \\
NGC 6181 &      1237662224092364842 &      247.8773876 &       20.093625 &     17.0 &    -15.7 &     0.25 &     24.0 &     198 & Y &     109.7 & SDSS \\
NGC 6181 &      1237662698115432783 &      248.1520797 &       19.810259 &     17.6 &    -15.1 &     0.29 &     23.4 &      37 & Y &     -88.2 & MMT \\
NGC 6181 &      1237662662147638034 &      248.5806385 &       19.720801 &     18.2 &    -14.5 &     0.86 &     24.3 &     285 & Y &      77.3 & AAT \\
NGC 6181 &      1237662224092496776 &      248.1953686 &       19.867013 &     18.6 &    -14.1 &     0.23 &     23.2 &      65 & Y &      95.3 & MMT \\
NGC 6181 &      1237662698115433445 &      248.1634268 &       19.792208 &     20.3 &    -12.4 &     0.19 &     23.5 &      47 & Y &    -135.3 & MMT \\\hline
NGC 5750 &      1237648721248845970 &      221.3161158 &       -0.159937 &     14.9 &    -17.1 &     0.41 &     21.6 &     105 & Y &      28.5 & SDSS \\\hline
NGC 7716 &      1237666408439939282 &      354.3506000 &        0.390803 &     13.7 &    -19.0 &     0.41 &     23.7 &     144 & Y &     102.5 & SDSS \\
NGC 7716 &      1237663277925204111 &      354.1952297 &        0.623424 &     15.6 &    -17.1 &     0.41 &     22.9 &     201 & Y &      88.9 & SDSS \\
NGC 7716 &      1237666408439677694\tablenotemark{1} &      353.7788053 &        0.301059 &     21.3 &    -11.4 &     1.39 &     23.2 &     213 & N &    -158.4 & MMT \\\hline
NGC 1015 &      1237678881574944900 &       39.9254643 &       -1.418742 &     17.0 &    -15.9 &     0.35 &     21.2 &     253 & Y &       5.9 & MMT \\
NGC 1015 &      1237678881574814166 &       39.5361613 &       -1.396696 &     20.1 &    -12.8 &     0.06 &     24.2 &      51 & Y &     -95.8 & AAT \\\hline
PGC068743 &      1237680192048857102 &      336.0481612 &       -3.482939 &     13.5 &    -19.5 &     0.41 &     22.5 &      98 & Y &       5.0 & SDSS \\
PGC068743 &      1237680066954264778 &      335.8363054 &       -3.659803 &     14.9 &    -18.1 &     0.56 &     21.8 &     164 & Y &     233.0 & SDSS \\
PGC068743 &      1237679996084617517 &      335.9799762 &       -3.270549 &     15.9 &    -17.1 &     0.58 &     21.4 &     119 & Y &      23.0 & AAT \\
PGC068743 &      1237680066954330699 &      335.9538495 &       -3.701195 &     19.4 &    -13.5 &     0.18 &     23.8 &     186 & Y &     -22.0 & AAT \\\hline
NGC 2543 &      1237657607497318756 &      123.2431732 &       36.198360 &     15.4 &    -17.5 &     0.30 &     21.5 &      37 & Y &      19.7 & SDSS \\
NGC 2543 &      1237657607497515484 &      123.6498976 &       36.434355 &     16.4 &    -16.5 &     0.35 &     23.0 &     246 & Y &      -8.0 & SDSS \\\hline
NGC 7541 &      1237679005021831220 &      348.6438163 &        4.498443 &     12.7 &    -20.1 &     0.68 &     20.1 &      34 & Y &     -15.0 & SDSS \\
NGC 7541 &      1237678777399443498 &      348.6966489 &        4.639955 &     15.0 &    -17.9 &     0.19 &     21.8 &      68 & Y &     175.8 & MMT \\
NGC 7541 &      1237678776862572690 &      348.7769076 &        4.373197 &     15.6 &    -17.2 &     0.55 &     20.8 &     120 & Y &     -19.7 & MMT \\
NGC 7541 &      1237679005021831801 &      348.6214885 &        4.507171 &     17.8 &    -15.0 &     0.55 &     23.5 &      43 & N &    -104.8 & AAT \\
NGC 7541 &      1237678777399509170 &      348.8741991 &        4.613261 &     19.9 &    -12.9 &     0.46 &     22.6 &     133 & Y &      17.8 & MMT \\
NGC 7541 &      1237679005558702536\tablenotemark{1} &      348.5545917 &        4.915003 &     20.7 &    -12.2 &     0.21 &     23.7 &     259 & Y &     119.4 & MMT \\ \hline
 \multicolumn{12}{c}{Incomplete Hosts}\\
NGC 4158 &      1237668298203267092 &      182.9906665 &       20.027849 &     14.4 &    -18.4 &     0.35 &     20.4 &     150 & Y &     -49.4 & SDSS \\
NGC 4158 &      1237668298203070473 &      182.4280443 &       20.046919 &     16.7 &    -16.1 &     0.37 &     23.5 &     231 & Y &      19.3 & MMT \\
NGC 4158 &      1237668298740007188 &      182.6898152 &       20.592928 &     18.2 &    -14.6 &     0.29 &     23.3 &     271 & Y &      79.9 & MMT \\
NGC 4158 &      1237668298203202132 &      182.8482068 &       20.063223 &     19.9 &    -12.9 &     0.07 &     23.2 &      78 & Y &    -115.3 & MMT \\\hline
NGC 3067 &      1237664338780029261\tablenotemark{1} &      149.6871689 &       32.720265 &     20.7 &    -11.3 &     0.42 &     24.6 &     158 & Y &      43.7 & MMT \\\hline
NGC 5792 &      1237648702984683605 &      225.0054012 &       -1.091302 &     14.8 &    -17.4 &     0.25 &     21.0 &     203 & Y &     -34.0 & SDSS \\
NGC 5792 &      1237655693015056396 &      224.5326175 &       -1.312596 &     15.3 &    -17.0 &     0.57 &     21.5 &     114 & Y &      24.5 & SDSS \\\hline
NGC 4030 &      1237650372092690464 &      180.2954422 &       -1.297684 &     13.4 &    -18.4 &     0.31 &     21.4 &     113 & Y &       1.0 & SDSS \\
NGC 4030 &      1237650762927308814 &      179.6917284 &       -1.461941 &     13.9 &    -17.9 &     0.40 &     21.1 &     220 & Y &      34.0 & SDSS \\\hline
NGC 7818 &      1237680247351738669 &        1.1296580 &       20.718018 &     15.8 &    -16.7 &     0.40 &     22.4 &      73 & Y &    -134.0 & SDSS \\\hline
NGC 3277 &      1237667287812735027 &      157.7782743 &       28.796645 &     12.9 &    -19.0 &     0.61 &     19.1 &     208 & Y &      14.6 & SDSS \\
NGC 3277 &      1237667255616143515 &      158.0432463 &       28.483057 &     15.2 &    -16.7 &     0.50 &     21.1 &      71 & Y &     173.9 & SDSS \\
NGC 3277 &      1237665367429677221 &      158.7123978 &       28.663853 &     15.5 &    -16.5 &     0.38 &     22.1 &     191 & Y &     -85.0 & SDSS \\
NGC 3277 &      1237667255616143540 &      158.0885195 &       28.419853 &     16.7 &    -15.2 &     0.47 &     22.6 &      66 & Y &     214.2 & SDSS \\
NGC 3277 &      1237667255616274560 &      158.3517683 &       28.552914 &     17.8 &    -14.1 &     0.29 &     22.3 &      48 & Y &    -122.5 & MMT \\\hline
UGC 00903 &      1237679169841725673 &       20.7768904 &       17.891450 &     17.1 &    -15.9 &     0.23 &     23.1 &     291 & Y &      23.3 & MMT \\
UGC 00903 &      1237678602387456130 &       20.2852690 &       17.602236 &     17.3 &    -15.7 &     0.40 &     23.0 &     105 & Y &     -53.7 & MMT \\
UGC 00903 &      1237678602387521791 &       20.5363427 &       17.528143 &     18.9 &    -14.0 &     0.34 &     23.2 &      70 & Y &      75.5 & MMT \\
UGC 00903 &      1237678602387456305 &       20.3284666 &       17.753939 &     19.6 &    -13.4 &     0.43 &     23.9 &     133 & Y &     -46.2 & MMT \\
\enddata
\tablenotetext{1}{These satellites are fainter than our survey completeness limit.}
\tablecomments{Column (1) indicated the satellite's host galaxy. Column (2)--(8) are the photometric properties of the satellite from the NASA-Sloan Atlas or SDSS DR12.  Column (9) is the satellite's project distant from the host in kpc.  Column (10) notes whether the object is star-forming based on the presence of H$\alpha$ in the discovery spectrum.  Column (11) is the velocity of the satellite minus the host velocity. Column (12) indicates the telescope which first obtained the satellite redshift (see \autoref{sssec_data} for telescope details).}
\end{deluxetable*}

\newpage

\newpage

\end{document}